\documentclass{article}

\PassOptionsToPackage{numbers}{natbib}

  \usepackage[eandd, final]{neurips_2026}

\usepackage[utf8]{inputenc} 
\usepackage[T1]{fontenc}    
\usepackage{hyperref}       
\usepackage{bookmark}       
\usepackage{graphicx}       
\usepackage[most]{tcolorbox} 
\usepackage{url}            
\usepackage{booktabs}       
\usepackage{amsmath}        
\usepackage{amsfonts}       
\usepackage{nicefrac}       
\usepackage{microtype}      
\usepackage{xcolor}         
\usepackage{enumitem}       
\usepackage{wrapfig}        
\usepackage{makecell}       
\usepackage{etoc}           
\usepackage{fontawesome5}   
\usepackage{pdfpages}

\newtcblisting{promptbox}[1]{
  listing only,
  breakable,
  enhanced,
  colback=gray!4,
  colframe=gray!55!black,
  coltitle=white,
  fonttitle=\bfseries\small,
  title={#1},
  left=4pt, right=4pt, top=4pt, bottom=4pt,
  boxrule=0.5pt,
  arc=2pt,
  listing options={
    basicstyle=\scriptsize\ttfamily,
    breaklines=true,
    breakatwhitespace=false,
    columns=fullflexible,
    keepspaces=true,
    showstringspaces=false,
    breakindent=0pt,
    postbreak=\mbox{\textcolor{gray}{$\hookrightarrow$}\space},
  }
}

\newtcolorbox[auto counter]{examplebox}[2][]{
  enhanced,
  breakable,
  colback=gray!3,
  colframe=black!55,
  boxrule=0.5pt,
  arc=2pt,
  left=6pt,right=6pt,top=4pt,bottom=4pt,
  fonttitle=\bfseries\small,
  coltitle=black,
  colbacktitle=gray!12,
  fontupper=\small,
  title={Example~\thetcbcounter: #2},
  #1
}

\usepackage{xspace}
\usepackage{pifont}
\newcommand{\method}{\textsc{P}\scalebox{0.9}{\textsc{rivacy}}\textsc{Sim}\xspace}
\newcommand{\ie}{i.e.,\xspace}
\newcommand{\eg}{e.g.,\xspace}
\newcommand{\xmark}{\ding{55}}

\title{\method{}: Evaluating LLM Simulation of \\ User Privacy Behavior}

\author{%
  James Flemings\\
  University of Southern California\\
  \texttt{jamesf17@usc.edu} \\
  \And
  Murali Annavaram \\
  University of Southern California\\
  \texttt{annavara@usc.edu} \\
}

\begin{document}

\maketitle

\begin{center}
  \href{https://huggingface.co/datasets/jamesflemings/PrivacySIM}{\raisebox{-0.18em}{\includegraphics[height=1em]{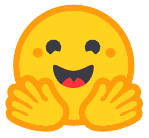}}\,Data}\quad
  \href{https://github.com/james-flemings/PrivacySIM}{\faGithub\,Code}
\end{center}

\etocdepthtag.toc{mainmatter}

\begin{abstract}
  Large language models (LLMs) are increasingly used to simulate human behavior, but their ability to simulate \emph{individual} privacy decisions is not well understood. In this paper, we evaluate whether a core set of user persona attributes can drive LLMs to simulate individual-level privacy behavior across various contexts. We introduce \method{}, an evaluation suite that benchmarks LLM simulation of user privacy behavior against the ground-truth responses of $1{,}000$ real users. These users are drawn from five published user studies on privacy behavior spanning LLM healthcare consultations, conversational agents, and chatbots. Drawing on these user studies and prior work in privacy behavior, we hypothesize three persona facets as plausible predictors of privacy decision-making: demographics, previous experiences, and stated privacy attitudes. We condition nine frontier LLMs on subsets of these three facets and measure how often each model's response to a data-sharing scenario matches the user's actual response. Our findings show that (1) zero-shot conditioning on these facets consistently improves simulation quality, but the performance of the best model (40.4\% accuracy) suggests that richer and more nuanced persona facets are needed to reliably align with users' mental models for privacy decisions. (2) A user's stated privacy attitudes alone are often not the best predictor because they can diverge from the user's actual privacy behavior. (3) Users with high AI/chatbot experience are the most challenging to simulate. \method{} is a first step toward understanding and improving the capabilities of LLMs to simulate user privacy decisions. We release \method{} to enable further evaluation of LLM privacy simulation.
\end{abstract}

\section{Introduction}
\label{sec:introduction}

Applying large language models (LLMs) to simulate users is a promising direction for evaluating the privacy of data-sharing systems~\citep{zhang2025searching} or serving as a proxy for human privacy preferences and norms. To illustrate, suppose a practitioner designing a healthcare assistant wants to know whether it is appropriate to share a user's mental-health information with family members, clinicians, or government agencies. Eliciting the user's data-sharing preferences through controlled user studies is the gold standard but is slow, expensive, and prone to privacy fatigue, where users make ineffective privacy decisions when feeling overwhelmed~\citep{acquisti2006there,choi2018role}. In contrast, a practitioner could condition an LLM to act as a user with a certain privacy concern level, \eg \texttt{"concerned with sharing their data"}, when using the healthcare assistant. After simulation, the privacy evaluation can measure the assistant's inappropriate data disclosures based on the simulated user's privacy behavior. This process reduces the need to elicit responses from human participants and enables quick and inexpensive privacy evaluation early in the design process~\citep{aher2023using,argyle2023out,park2023generative}. The more costly, but accurate, human studies can then validate the system's privacy closer to deployment.

However, it remains an open question whether LLMs as simulators can reflect individual privacy behavior accurately enough to be useful for these purposes. In other words, do the actions from an LLM simulating a user \texttt{"concerned with sharing their data"} truly reflect a concerned user's data-sharing actions? Existing work simulating user behavior typically conditions LLMs on personas, representations of hypothetical individuals~\citep{hwang2023aligning,joshi2025improving,venkit2026need}. Yet the insights derived from these works may not transfer to simulating \emph{privacy behavior} of a user. Users routinely deviate from their stated attitudes when making privacy decisions, a pattern known as the \emph{privacy paradox}~\citep{colnago2023there, kokolakis2017privacy,norberg2007privacy}. For example, a user concerned about disclosing personal information to an LLM may still share it to receive personalized responses\cite{mireshghallah2024trust, zhang2024s}. Therefore, simulating privacy behavior requires going beyond generalized user personas to \emph{privacy personas} that are tailored to user privacy. 

\begin{figure}[t]
  \centering
  \includegraphics[width=\linewidth]{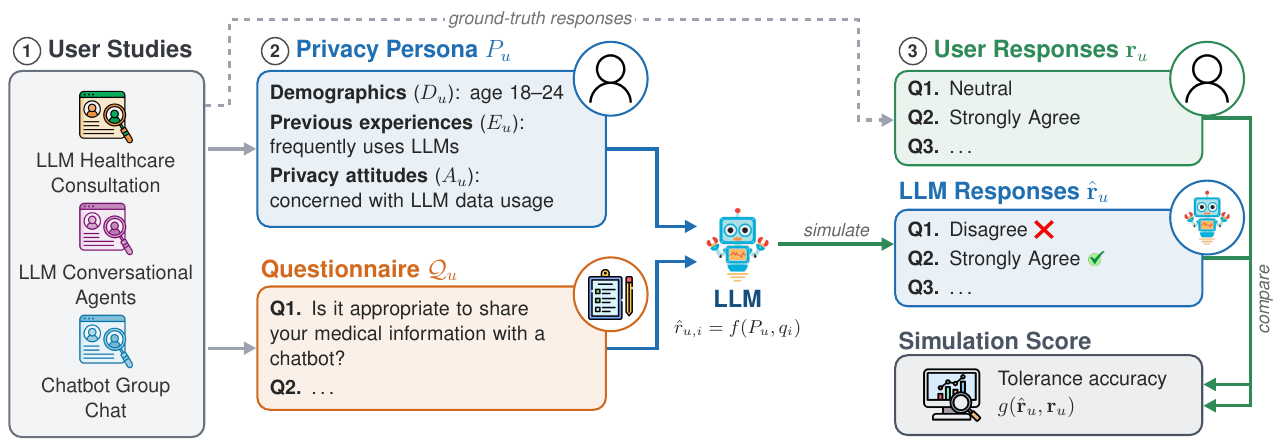}
  \caption{Overview of \method{}, an evaluation suite for simulating user privacy behavior. We collect user responses and questionnaires from existing user studies on privacy behavior in AI and chatbot contexts. We then condition LLMs on subsets of a user's privacy persona (demographics, previous experiences with AI and chatbots, and stated privacy attitudes) to simulate their responses to data-sharing questions. Finally, we evaluate the simulation quality by comparing the LLM's simulated responses to the user's actual responses.}
  \label{fig:framework}
\end{figure}

Our work seeks to \emph{identify a core set of persona facets that can be used by an LLM to accurately simulate individual-level privacy behavior.} Such a set would allow practitioners to condition LLMs on these privacy-driven facets to simulate users across various contexts. Despite the practical promise, to our knowledge no standardized evaluation suite or methodology exists for assessing the fidelity of LLMs simulating \emph{individual} privacy behavior. The closest recent work either evaluates only population-level alignment~\citep{li2026well} or uses prior privacy decisions to predict subsequent ones~\citep{fawaz2026text}. In contrast, we target individual-level alignment using a core set of persona facets, without assuming access to in-domain prior user privacy decisions.

We address this gap by introducing \method{} (shown in Figure~\ref{fig:framework}). At the core is a curated dataset of users' responses to a set of data-sharing scenarios posed across five existing user studies. These five studies were conducted in diverse areas such as LLM healthcare consultation, conversational agents, and group chatbots. In particular, we sample 200 users from each study and 7 to 10 data-sharing questions per user, yielding 9,400 user--question pairs. From this curated dataset, we hypothesize three core facets that together constitute a user's \emph{privacy persona}: demographics (\eg age), previous experiences (\eg AI or chatbot usage), and privacy attitudes (\eg general data-usage concerns). Crucially, \method{} is grounded in real user responses to data-sharing scenarios with a rich set of user attributes, enabling accurate evaluation of whether LLMs can faithfully simulate the nuanced, often inconsistent ways individual users handle privacy~\citep{acquisti2005privacy}. We condition nine frontier LLMs on subsets of these facets and measure how closely their simulated responses match the users' responses.

We use \method{} to organize our analysis around three research questions (RQs):
\begin{itemize}[nosep,leftmargin=1.2em]
    \item \textbf{RQ1.} To what extent can a core set of persona facets be used by LLMs to simulate user privacy decisions across data-sharing scenarios?
    \item \textbf{RQ2.} How does LLM simulation quality vary across user privacy persona groups (\eg users with high stated concern and high AI/chatbot exposure vs.\ users with low stated concern and high AI/chatbot exposure)?
    \item \textbf{RQ3.} How do LLM characteristics (model size and reasoning effort) and privacy-informed prompts affect simulation quality?
\end{itemize}

Our key findings are as follows. First, while demographics, previous experiences, and privacy attitudes consistently improve simulation quality, the top-scoring LLM, Gemini 3.1 Pro, achieves only 40.4\% accuracy. Contrary to expectation, privacy attitudes do not provide the strongest signal of a user's actual privacy behavior. This finding highlights the challenge of the privacy paradox when simulating human privacy behavior (\textbf{RQ1}). Second, we group users along two axes, stated privacy concern and AI/chatbot familiarity, inspired by \citet{dupree2016privacy}. Within this grouping, users with high AI/chatbot experience have the lowest simulation quality, indicating that certain user privacy groups are disproportionately difficult to model (\textbf{RQ2}). Third, scaling model size and reasoning effort, as well as prompting models to reason through well-established privacy theories, such as Privacy Calculus~\citep{culnan1999information}, yield negligible gains (\textbf{RQ3}). Taken together, these results point to eliciting richer user facets in order to reliably align LLMs with users' mental models for privacy decisions.
\section{Related Work}\label{sec:related_works}

\textbf{Privacy-decision theoretical frameworks.} Privacy theories provide a framework for how individuals make data-sharing decisions. Contextual Integrity~\citep{nissenbaum2004privacy} posits that disclosure is appropriate when it conforms to the privacy norms that are contextually relevant the originating context. Privacy Calculus~\citep{culnan1999information} models disclosure as a deliberate cost-benefit trade-off between perceived benefits and risks, while Bounded Rationality accounts~\citep{acquisti2004privacy,acquisti2005privacy} argue that users instead rely on heuristics and satisficing. Protection Motivation Theory~\citep{rogers1975protection}, originally developed for health-behavior research, casts protective behavior as the joint output of a threat appraisal and a coping appraisal. 

\textbf{User studies on privacy behavior.} Privacy theories may not always capture real-world nuances of human privacy behavior. To provide a more on-the-ground view, user studies elicit responses to various data-sharing contexts with privacy implications, such as app permissions~\citep{felt2012android, wijesekera2017feasibility}, smart home devices~\citep{abdi2021privacy,apthorpe2018discovering}, and social media platforms~\citep{wang2011regretted}. Recent attention has turned to user privacy behavior in data-sharing contexts involving LLMs and chatbots, the setting we study. Prior work has studied disclosure of information to LLMs~\citep{malki2025hoovered, zhang2024s, zufferey2025ai}, disclosure of LLM chat history to other entities~\citep{liu2025prevalence, tran2025understanding}, and user data access by chatbots and AI agents~\citep{chou2026bot,wu2025towards}. These studies usually pose data-sharing contexts as vignette-based questionnaires~\citep{martin2016measuring} grounded in Contextual Integrity. Lastly, prior work has clustered users into privacy-persona groups based on attitudes and behaviors~\citep{dupree2016privacy, kumaraguru2005privacy}. 

\textbf{LLM simulation of user behavior.} An emerging and promising direction is to use LLMs to simulate user behavior instead of eliciting preferences from users~\citep{aher2023using,argyle2023out,bisbee2024synthetic,dominguezolmedo2024questioning,santurkar2023whose,hu2025simbench}. A common instantiation conditions LLMs on synthesized personas from seed human data~\citep{hwang2023aligning, joshi2025improving, park2023generative,park2024generative,venkit2026need}. Most recently, two studies investigate using LLMs conditioned on personas to simulate user privacy behavior. \citet{li2026well} evaluate population-level alignment of LLMs to simulate privacy attitudes and behavior. They generate synthetic participants by sampling demographics that match the distribution reported in the human studies they replicate. \citet{fawaz2026text} generate user personas by synthesizing in-domain prior user privacy decisions to predict new ones~\citep{flemings2025personalizing}. In contrast to both, our work focuses on individual-level alignment using a core set of privacy persona attributes, without assuming access to in-domain prior user privacy decisions.
\section{Constructing \method{}}
\label{sec:methodology}

\method{} grounds LLM simulation of privacy behavior in real user responses from five published user studies, conditions models on a user's privacy persona, and scores each simulated decision against the user's decision. We formalize the simulation task in Section~\ref{sec:problem-statement}, describe how we curate the evaluation suite in Section~\ref{sec:dataset-construction}, and define our evaluation metric in Section~\ref{sec:evaluation-methodology}.

\subsection{Problem Statement}
\label{sec:problem-statement}

Let $P_u$ denote the \emph{privacy persona} of a user $u$, which plausibly captures signal about the user's privacy behavior (see Section~\ref{sec:dataset-construction} for a concrete example). Let $\mathcal{Q}_u$ be the questionnaire of data-sharing scenarios given to user $u$, and let $q_i \in \mathcal{Q}_u$ be the $i$-th question, paired with the user's response $r_{u,i}$. An LLM $f$ takes the persona and the question and produces a simulated response: $\hat{r}_{u,i} \;=\; f(P_u, q_i).$ Define $\mathbf{r}_{u} = [r_{u,1}, r_{u,2}, \ldots, r_{u,|\mathcal{Q}_u|}]$ and $\hat{\mathbf{r}}_{u} = [\hat{r}_{u,1}, \hat{r}_{u,2}, \ldots, \hat{r}_{u,|\mathcal{Q}_u|}]$ to be the user's and LLM's complete responses to the questionnaire $\mathcal{Q}_u$. We introduce an evaluation function $g(\hat{\mathbf{r}}_{u}, \mathbf{r}_{u})$ to measure the alignment of the LLM's simulated responses with the user's responses. We note that neither the data-sharing questions in $\mathcal{Q}_u$ nor the corresponding user responses appear in $P_u$.

\subsection{Dataset Curation}
\label{sec:dataset-construction}
As mentioned in Section~\ref{sec:introduction}, no standardized evaluation suite exists for assessing LLM simulation of user privacy behavior. While one option could be to synthetically generate users using an LLM, such an approach risks missing the complexity and nuances inherent in real individuals' mental models for privacy decision-making~\citep{kang2015my}. Hence, we curate a dataset grounded in existing user studies on privacy behavior. 

To collect a set of questionnaires $\mathcal{Q}_{u}$ and personas $P_u$ to be used in \method{}, we search for prior work that meets two main criteria. The first is that each study contains user responses to questions that can be separated into user privacy personas $P_u$ and data-sharing questionnaire $\mathcal{Q}_u$, which allows us to condition the LLM on these personas and evaluate its ability to simulate the user's responses to data-sharing scenarios. The second criterion is that we restrict the data-sharing scenarios $\mathcal{Q}_u$ to involve AI, chatbots, or agents. We make this choice for two reasons. The first is that privacy norms and preferences are still emerging in this space, so LLMs are unlikely to have this knowledge, which adds to the challenge of the simulation. In contrast, privacy norms and preferences are well studied in established contexts such as social media, mobile apps, and Internet-of-Things devices. The second is that, given the recently heightened concern about user privacy in AI interactions, this exploration addresses the timely question of whether LLMs can understand user data-sharing decisions when AI or chatbots are involved. 

\begin{table}[t]
\centering
\caption{Summary of source user studies in \method{}. $D_u$, $E_u$, and $A_u$ denote demographics, previous experiences, and privacy attitudes, respectively. ``Domain'' refers to the data-sharing context posed in the original study.}
\label{tab:datasets}
\small
\resizebox{\textwidth}{!}{%
\begin{tabular}{@{}llllccll@{}}
\toprule
Author & Venue & Country & $D_u$ & $E_u$ & $A_u$ & Domain & Response scale\\
\midrule
\citet{liu2025prevalence}      & S\&P 2025  & China   & \xmark      & \checkmark & \checkmark & LLM Healthcare Consultation & 5-point\\
\citet{tran2025understanding} & AIES 2025  & US      & \xmark      & \checkmark & \checkmark & LLM Chatbot & 0--100 \\
\citet{wu2025towards}           & S\&P 2026 & US      & \checkmark & \checkmark & \checkmark & AI Agent Permissions & 4-point\\
\citet{zufferey2025ai}    & PETS 2025  & Various & \checkmark & \checkmark & \checkmark & LLM Conversational Agents & 5-point \\
\citet{chou2026bot}           & PETS 2026  & Various & \checkmark & \checkmark & \checkmark & Chatbot Group Chat & 5-point \\
\bottomrule
\end{tabular}%
}
\end{table}

Finding user studies that meet these criteria is challenging because many are not publicly available. Eventually, we found five user studies that meet these criteria, summarized in Table~\ref{tab:datasets}. Notably, each study is published in a top-tier privacy venue, and participants span multiple countries.  Each study varies in the type of questions asked and the response format used. Analyzing the responses from each of the five user studies, we derive \emph{three persona facets to instantiate a user's privacy persona $P_{u}$ that we hypothesize to be plausible predictors of privacy behavior}:
\begin{itemize}[nosep,leftmargin=1.2em]
    \item \textbf{Demographics ($D_u$)}: basic user information such as age, gender, and education.
    \item \textbf{Previous Experiences ($E_u$)}: the user's familiarity with and knowledge of emerging technologies such as LLM usage frequency.
    \item \textbf{Privacy Attitudes ($A_u$)}: the user's general privacy attitudes and concerns regarding data practices, \eg the Internet Users' Information Privacy Concerns (IUIPC) scale~\citep{malhotra2004internet}.
\end{itemize}

Privacy attitudes and concerns are a natural choice for our evaluation because they are widely studied in privacy theories and user studies for their influence on privacy decision-making~\citep{baruh2017online, hargittai2010facebook}. However, a recurring finding across these works is the \emph{privacy paradox}, the gap between users' stated privacy attitudes and their actual data-sharing behavior, motivating us to investigate additional persona facets. One such persona facet is a user's knowledge in or previous experiences with emerging technology, which has also been linked to user privacy behavior~\citep{hargittai2010facebook, zeng2017end}. Lastly, demographics are a common attribute for conditioning LLMs in prior simulation work~\citep{venkit2026need}, so we also measure their effectiveness in simulating privacy behavior. All five user studies have released user responses to questions that can fall under previous experiences and privacy attitudes (demographics are optional because some publicly released user studies omit them for confidentiality), but we could not incorporate additional facets that are available across all five studies (more discussion on this is in Section~\ref{sec:discussion}). 

We keep each questionnaire presented to the LLM as close as possible to the original questionnaire from the source study. This minimizes variation in how the questions are posed to the original study participants and to the LLM. Example~\ref{ex:user-example} shows a representative user from \citet{zufferey2025ai} with one attribute drawn from each of the three persona facets and one paired data-sharing question and response (an example user for each user study can be found in Appendix~\ref{appendix:user_studies}). Full prompts containing a complete user privacy persona followed by the questionnaire for each study can be found in in Appendix~\ref{app:prompts}. To ensure each user study has equal weight in the final evaluation, we sample 200 users per study, since the smallest user study has 203 users; for each user we sample ten data-sharing questions in four of the studies and use all seven data-sharing questions in the Chatbot Group Chat user study, yielding a benchmark of 1,000 users and 9,400 user--question pairs. Lastly, all datasets are publicly released by the original authors under permissive licenses (Creative Commons or MIT) that allow reuse if the original authors are properly attributed. Appendix~\ref{appendix:user_studies} gives more information about each user study.

\begin{examplebox}[label=ex:user-example]{User from LLM Conversational Agents \citep{zufferey2025ai}}
\textbf{Demographics ($D_u$).}\quad Age: \emph{18--29}.\\[2pt]
\textbf{Previous experiences ($E_u$).}\quad \emph{Q.} Are you using your own local chatbot instance (\eg Llama, Mistral) based on only locally stored data?\quad \emph{A.} ``Yes, but not exclusively (I also use chatbot services such as ChatGPT).''\\[2pt]
\textbf{Privacy attitudes ($A_u$).}\quad \emph{Q.} It usually bothers me when online companies ask me for personal information.\quad \emph{A.} ``Agree'' (7-point: Strongly disagree~$\to$~Strongly agree).

\vspace{2pt}
\hrule height 0.3pt
\vspace{2pt}

\textbf{Data-sharing question $q_i$ (Q1).}\quad How frequently do you share \emph{personal identifiers} (\eg name, ID card number, email, phone number) with a chatbot?\\[2pt]
\textbf{User response $r_{u,i}$.}\quad ``I often share'' \;(5-point: \emph{never / rarely / sometimes / often / usually shared}; mapped to $\{0,1,2,3,4\}$ for evaluation).
\end{examplebox}

\subsection{Evaluation Methodology}
\label{sec:evaluation-methodology}
Because we preserve the original questionnaires, the response scales differ across studies. For example, responses to questions from \citet{zufferey2025ai} are frequency disclosures on a 5-point Likert scale (\ie $r_{u,i} \in \{0,1,2,3,4\}$), whereas \citet{tran2025understanding} elicit agreement with privacy statements on a 0--100 numeric scale. Hence, we cannot simply use exact-match accuracy here because responses on the 0--100 scale rarely match exactly. 

To address this shortcoming, we modify exact-match accuracy so that it measures whether the LLM's simulated response falls within a certain range of the user's actual response. This bound is defined as a fraction of the response scale's range. More precisely, let $W_{\mathcal{Q}_u}$ be the upper bound of the response scale used in $\mathcal{Q}_u$ (\eg $4$ for a 5-point Likert scale, $100$ for a 0--100 numeric scale), so each response lies in $[0, W_{\mathcal{Q}_u}]$. The tolerance accuracy of the simulation for a user $u$ with responses $\mathbf{r}_u$ and simulated responses $\hat{\mathbf{r}}_u$ is
\vspace{-6pt}
\begin{equation}\label{eq:tau_accuracy}
g(\hat{\mathbf{r}}_{u}, \mathbf{r}_{u}) \;=\; \frac{1}{|\mathcal{Q}_u|} \sum_{i=1}^{|\mathcal{Q}_u|} \mathbf{1}\!\left[\,\lvert r_{u,i} - \hat{r}_{u,i}\rvert \;\leq\; \tau \cdot W_{\mathcal{Q}_{u}}\,\right].
\end{equation}
\vspace{-6pt}

We set $\tau=0.15$ because it addresses the above shortcoming while reducing to exact match on the 4-point and 5-point Likert scales. On the 0--100 scale, the LLM is credited with a match when its prediction lies within $\pm 15$ points of the user's response. We performed a sweep over $\tau$ in Appendix~\ref{appendix:tau_sweep}, showing that $\tau=0.15$ is an appropriate choice. We report $100 \cdot g(\hat{\mathbf{r}}_{u}, \mathbf{r}_{u})$ as a percentage in our experimental results.

\textbf{Evaluation use case.} LLM evaluations increasingly rely on synthetic datasets. However, we argue that evaluating LLM privacy decision-making should not rely on synthetic datasets, as such datasets may not fully capture the complexity and nuances of real-world privacy behavior. Hence, it is crucial that privacy evaluations be grounded in real user data, and \method{} fulfills this need by curating a dataset of real user responses. 

While our study focuses primarily on zero-shot conditioning of LLMs to test their innate ability to simulate privacy behavior, as this is a core promise of LLMs as simulators, we note that \method{} can be extended for few-shot in-context learning and fine-tuning evaluations. This setup shifts the research focus from a model's intrinsic simulation abilities to its capacity to learn privacy prediction tasks from training data (we include preliminary results of this setup with a non-LLM majority-class predictor in Appendix~\ref{appendix:majority_baseline}). Hence, \method{} enables the community to explore novel research dimensions in LLM simulation of user privacy behavior.
\section{Experiments}
\label{sec:experiments}

We now use \method{} to answer the three research questions (RQs) posed in Section~\ref{sec:introduction}. Section~\ref{sec:experimental-setup} describes our experimental setup, and Sections~\ref{sec:rq1}, \ref{sec:rq2}, and \ref{sec:rq3} present results for RQ1, RQ2, and RQ3, respectively.

\subsection{Experimental Setup}
\label{sec:experimental-setup}

\textbf{Models.}
We evaluate nine frontier reasoning LLMs, both open-weight and closed-source. The closed models are GPT-5.4 (\texttt{gpt-5.4-2026-03-05}), GPT-5.4 Mini (\texttt{gpt-5.4-mini-2026-03-17}), Claude Sonnet 4.6 (\texttt{claude-sonnet-4-6}), Gemini 3.1 Pro (\texttt{gemini-3.1-pro-preview}), and Gemini 3 Flash (\texttt{gemini-3-flash-preview}), accessed through their respective vendor APIs. The open-weight models are Qwen3.5 27B and Qwen3.5 122B~\citep{qwen3.5}, Nemotron 3 120B~\citep{nvidia_nemotron_3_2025}, and Gemma 4 31B~\citep{team2026gemma}, served locally with vLLM~\citep{kwon2023efficient} on an internal cluster of $8\,{\times}$\,A100 (40\,GB) GPUs.

\textbf{Decoding parameters.}
For open-weight models we adopt the temperature $T$ and top-$p$ values recommended by each model provider, namely $T=1.0$ and $p=0.95$\footnote{\url{https://huggingface.co/google/gemma-4-31B-it\#1-sampling-parameters}}\footnote{\url{https://huggingface.co/Qwen/Qwen3.5-27B\#using-qwen35-via-the-chat-completions-api}}.. For closed-source models, the temperature parameter is not configurable. Unless stated otherwise, we set the reasoning effort of each model to medium if applicable (Section~\ref{sec:rq3} varies this choice).

\textbf{Reporting conventions.}
Because a subset of the user studies does not contain demographics, we split the results into two groups: (1) \emph{5 studies} containing all the user studies and (2) \emph{3 studies} restricted to the domains AI Agent Permissions, Chatbot Group Chat, and LLM Conversational Agents, which all contain demographics. Unless stated otherwise, claims and overall trends are based on the 5 studies, while analyses involving demographics are based on the 3 studies. We assess the significance of certain experimental observations with Wilcoxon signed-rank tests on per-user simulation accuracies.

\begin{figure}[t]
  \centering
  \includegraphics[width=\linewidth]{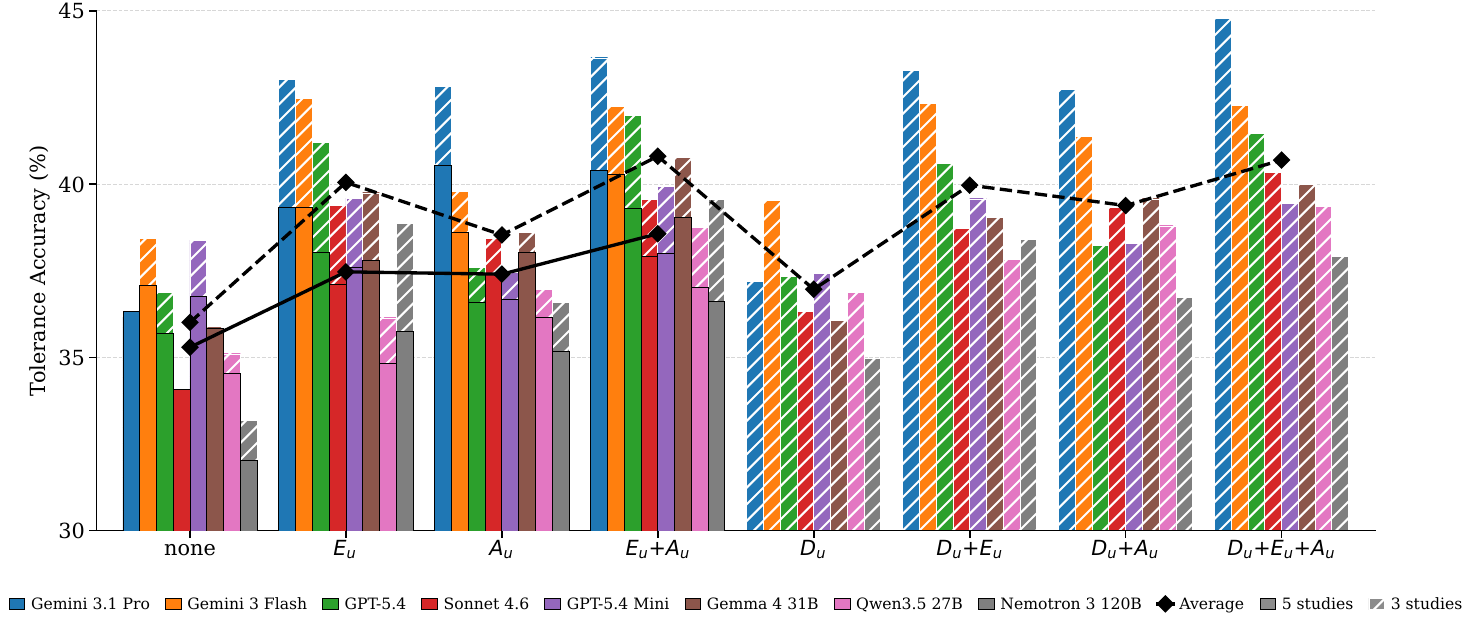}
  \caption{Average accuracy by prompt type and model across user studies. Solid bars are the average of all five user studies, and hatched bars are the average of the three user studies that contain demographics. The black diamond lines are the averages across the eight models. Our key findings show that (1) while previous experiences, privacy attitudes, and demographics consistently improve LLM simulation of user privacy behavior, the best performing model, Gemini 3.1, Pro achieves only 40.4\% on $E_u+A_u$ across the five studies. (2) Prompting without a privacy persona achieves accuracy above random choice, suggesting that LLMs posses and utilize knowledge of AI privacy norms and generally acceptable privacy behavior during simulation. (3) Privacy attitudes on average are not the best predictor of privacy behavior.}
  \label{fig:main_results}
\end{figure}

\subsection{RQ1: To what extent can privacy persona attributes drive accurate user simulations?}\label{sec:rq1}

\begin{figure}[t]
  \centering
  \includegraphics[width=\linewidth]{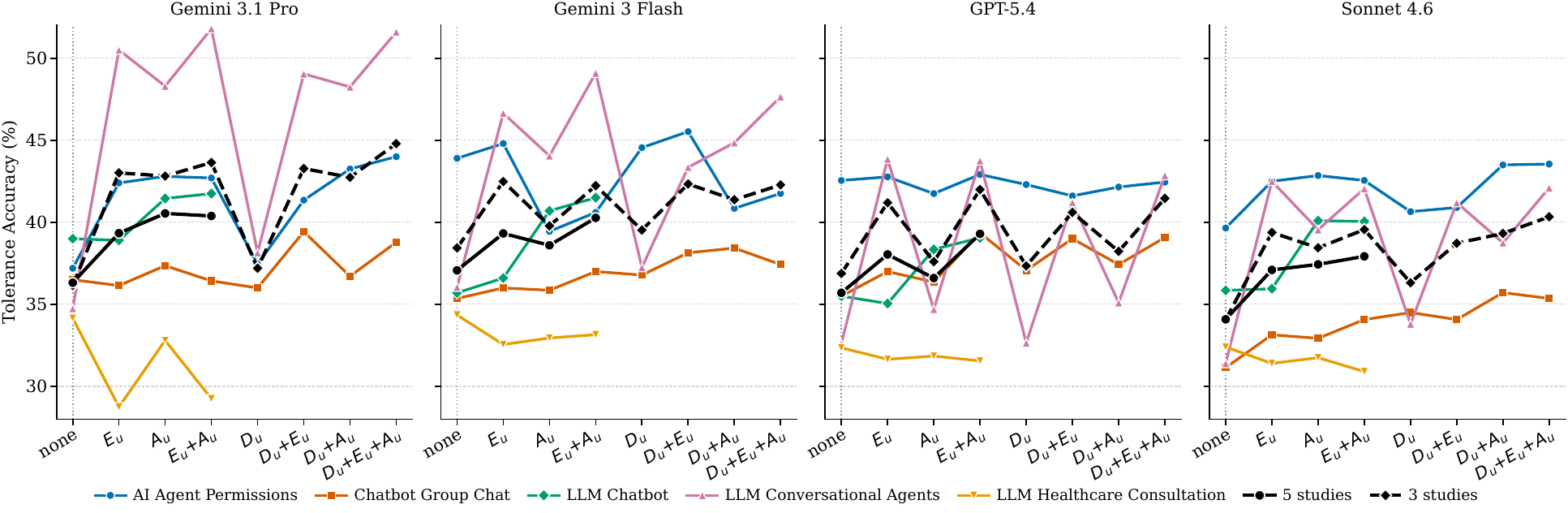}
  \caption{Per-domain accuracy with two averages (solid black: 5-study average; dashed black: 3-study average) across the eight prompt types for four closed-source models: Gemini 3.1 Pro, Gemini 3 Flash, GPT-5.4 (medium), and Claude Sonnet 4.6. We observe that the simulation quality can vary based on the domain of the data-sharing scenario. In particular, the simulation quality for LLM healthcare consultation is substantially lower compared to LLM conversational agents or chatbot group chat.}
  \label{fig:domain_subplots}
\end{figure}

To answer RQ1, we condition each LLM on every non-empty subset of the three persona facets ($\{D_u, E_u, A_u\}$) and without a privacy persona  (\texttt{none}), yielding eight prompt types. Figure~\ref{fig:main_results} reports the average accuracy of each model across prompt types. Figure~\ref{fig:domain_subplots} provides a per-model breakdown for four closed-source models, plotting each prompt type alongside each domain to analyze the cross-domain variance. From these results, we make the following observations.

\textbf{1. Demographics, previous experiences, and privacy attitudes provide modest improvement in simulating user privacy.}
Figure~\ref{fig:main_results} shows that conditioning on previous experiences and privacy attitudes ($E_u{+}A_u$) yields an average accuracy of $38.6\%$ across the eight models on the 5 studies, a $3.3\%$ increase over conditioning without a privacy persona ($p<0.001$). On the 3 studies that include demographics, conditioning on all three facets ($D_u{+}E_u{+}A_u$) yields $40.7\%$, a $4.7\%$ increase over no persona ($p<0.001$). These results suggest that LLMs can extract real signal from these three persona facets, but remain far from faithfully simulating individual privacy decisions, motivating further work on finding richer persona representations.

\textbf{2. LLMs conditioned without a privacy persona default to generally acceptable privacy behavior.}
Our results show that even without conditioning on any user information, LLMs still perform better than random chance, suggesting LLMs exhibit a non-trivial understanding of generally acceptable privacy behavior involving AI/chatbots. This finding is consistent with prior work evaluating whether LLMs can reason about contextual privacy, which shows that LLMs possess some awareness of privacy norms~\citep{mireshghallahcan,shao2024privacylens}. When analyzing the reasoning traces produced without persona facets, we find that LLMs usually assume an average user's privacy behavior and then apply population-level privacy norms relevant to the data-sharing question (we provide an example of GPT-5.4's reasoning trace for no-persona conditioning in Appendix~\ref{appendix:reasoning_trace}). Hence, the persona facets provide a more personalized privacy signal on top of an assumed general user.

\textbf{3. Privacy attitudes are vulnerable to the privacy paradox.}
In many of the models and domains (Figures~\ref{fig:main_results} and~\ref{fig:domain_subplots}), we observe a consistent dip in accuracy when conditioning on privacy attitudes ($A_u$) alone compared to conditioning on previous experiences ($E_u$) alone. We would expect that privacy attitudes would consistently provide the strongest signal because they are seemingly the most relevant to privacy behavior. We attribute this counterintuitive finding to the privacy paradox, where a user's stated privacy attitudes may not align with their actual privacy behavior, which was originally observed in some of our evaluated user studies~\citep{liu2025prevalence,zufferey2025ai}. This paradox makes it difficult for LLMs conditioned solely on self-reported attitudes to simulate user decisions accurately. However, when paired with previous experiences, the model can match and even slightly exceed the performance of previous experiences alone, suggesting that the two facets together can provide a more complete picture of user privacy behavior than either alone.

\textbf{4. Demographics alone yield the smallest single-facet gain.}
Our results show that conditioning on user demographics negligibly improves alignment with user privacy decisions. In other words, user demographics on their own provide little predictive signal for privacy behavior, a finding consistent with prior privacy user studies in which demographics did not explain users' privacy settings or preferences~\citep{hargittai2010facebook, lin2014modeling}. This finding contrasts with prior findings in LLM simulation where demographics generally provide useful predictive signal~\citep{venkit2026need}.

\textbf{5. Privacy behavior is easier to simulate in some domains than in others.}
In Figure~\ref{fig:domain_subplots}, we observe varying simulation performance depending on the domain. For example, all four models perform worst on LLM Healthcare Consultation under every prompt type. This may be because the healthcare domain involves more complex and high-stakes privacy decisions that are less predictable, whereas LLM Conversational Agents and Chatbot Group Chat domains may involve more consistent and norm-driven privacy behaviors that are easier for LLMs to simulate.

\subsection{RQ2: How does LLM simulation quality vary across user privacy persona groups?}
\label{sec:rq2}
Next, we are interested in understanding LLM simulation quality across different groups of users based on their privacy persona. To this end, we group users based on their stated previous experiences and privacy attitudes using a privacy persona taxonomy inspired by \citet{dupree2016privacy} and analyze the simulation quality across these groups. We detail below the persona grouping and per-cluster results.

\textbf{Persona grouping.}
We group users along two axes. \emph{Stance} is the user's stated privacy concern, computed as the average of their privacy-attitude responses ($A_u$), and \emph{exposure} is the user's self-reported familiarity with AI tools, computed as the average of their previous-experience responses ($E_u$). We standardize both scores within each study and assign each user to the nearest of five cluster centers: (1) high-stance / high-exposure, (2) high-stance / mid-exposure, (3) mid-stance / mid-exposure, (4) low-stance / high-exposure, and (5) low-stance / low-exposure. Full details are in Appendix~\ref{app:clustering}.

Table~\ref{tab:per-cluster} reports the accuracy at no persona (\texttt{none}) and $E_u{+}A_u$ per cluster, averaged over four closed-source models (Gemini 3.1 Pro, Gemini 3 Flash, GPT-5.4, and Claude Sonnet 4.6) on the 5 studies. From these results, we make the following observations:

\textbf{1. Mid-stance / mid-exposure users have the highest simulation quality when conditioned on no privacy persona.}
Users within this group represent the average user, and this observation serves as a good sanity check. As observed in Section~\ref{sec:rq1}, LLMs conditioned on no privacy persona assume an average user's privacy behavior, so they should align most closely with this group.

\textbf{2. High-exposure users are the hardest group to simulate when conditioned on a privacy persona.}
Users within this group are power users who either use AI frequently or use many features of AI agents and chatbots. These users are simulated with the lowest accuracy. We hypothesize that LLMs interpret high knowledge or experiences with AI as a sign that the user is more willing to share their data with AI. But in actuality, these users may be more restrictive than the LLM assumes.

\textbf{3. Mid-exposure users have the highest simulation accuracy with personas.}
In contrast, the high-stance / mid-exposure and mid-stance / mid-exposure clusters reach the highest and second-highest $E_u{+}A_u$ accuracy in the table, and the former also has the largest gain from persona conditioning ($+8.53\%$). We hypothesize that since these users are a tier below power users, their stated attitudes are consistent with their privacy behavior, allowing for more accurate inferences about their privacy decisions.

\textbf{4. Privacy personas are crucial for simulating high-stance users.}
When conditioning on no privacy persona, the high-stance / high-exposure and high-stance / mid-exposure clusters have the lowest accuracy. This shortcoming arises because the LLM falls back to an ``average user'' when not conditioned on any persona attributes, performing poorly on users whose actual privacy behavior is more restrictive. However, when conditioning on privacy personas, the high-stance / high-exposure and high-stance / mid-exposure clusters achieve the second-highest and highest accuracy gain, respectively. This improvement suggests that the most privacy-concerned users benefit the most from privacy personas.

\begin{table}[t]
  \begin{minipage}[t]{0.50\linewidth}
  \centering
  \caption{Accuracy for each privacy persona group averaged over Gemini 3.1 Pro, Gemini 3 Flash, GPT-5.4, and Claude Sonnet 4.6 on the 5 studies. Cluster labels are (stance / exposure) where stance is based on $A_u$ and exposure is based on $E_u$. $n$ is the number of users in each cluster. \textbf{Bold} marks the highest and lowest values of acc($E_u{+}A_u$).}
  \label{tab:per-cluster}
  \resizebox{\linewidth}{!}{%
  \begin{tabular}{p{2.4cm}rrr}
    \toprule
    Cluster (stance / exposure) & $n$ & acc(\texttt{none}) & acc($E_u{+}A_u$) \\
    \midrule
    high / high  & 161 & 33.90 & 37.25 \\
    high / mid   & 226 & 33.88 & \textbf{42.41} \\
    mid / mid    & 187 & 39.94 & 41.53 \\
    low / high   & 221 & 35.44 & \textbf{36.51} \\
    low / low    & 205 & 36.00 & 39.26 \\
    \bottomrule
  \end{tabular}}
\end{minipage}
\hfill
  \begin{minipage}[t]{0.46\linewidth}
  \centering
  \caption{Simulation accuracy across model families and variants, averaged over the 5 studies. For GPT-5.4 we report the smaller \emph{mini} model and the full model under \emph{low} and \emph{medium} reasoning effort. We find that reasoning effort and model size have negligible impact.}
  \label{tab:model-characteristics}
  \resizebox{\linewidth}{!}{%
  \begin{tabular}{llrr}
    \toprule
    Family & Variant & acc(\texttt{none}) & acc($E_u{+}A_u$) \\
    \midrule
    GPT-5.4  & Mini    & 36.75 & 37.99 \\
    GPT-5.4  & Low     & 36.14 & 38.77 \\
    GPT-5.4  & Medium  & 35.70 & 39.30 \\
    Gemini 3 & Flash   & 37.07 & 40.27 \\
    Gemini 3.1 & Pro   & 36.32 & 40.39 \\
    Qwen3.5  & 27B     & 34.54 & 37.03 \\
    Qwen3.5  & 122B    & 35.49 & 37.34 \\
    \bottomrule
  \end{tabular}}
\end{minipage}

\end{table}

\subsection{RQ3: How do LLM characteristics and privacy-informed prompts affect simulation?}
\label{sec:rq3}
Finally, we are interested in understanding how certain model characteristics (such as model size and reasoning effort) and prompts informed by established privacy theories affect simulation quality. All numbers in this section focus on accuracy across all 5 studies.

\textbf{Reasoning effort and model size have negligible effect.} From Table~\ref{tab:model-characteristics}, we observe that going from low to medium reasoning on GPT-5.4 changes the accuracy by less than $1\%$ ($p\geq0.16$). This result suggests that simulating user privacy decisions does not benefit from scaling test-time compute. Moreover, at $E_u{+}A_u$, we observe negligible gains in simulation quality when scaling the model size for GPT-5.4, Gemini 3, and Qwen 3.5.  

\setlength{\intextsep}{0pt}
\begin{wraptable}{r}{0.5\textwidth}
  \centering
  \caption{Privacy-informed prompts, based on Privacy Calculus, Bounded Rationality, and Protection Motivation Theory (PMT), compared to the standard persona prompt on Gemini 3 Flash and Gemma 4 31B. Numbers are average accuracy across the 5 studies. We restrict to prompt types that exclude demographics so all 5 studies are comparable. We find that no privacy-informed prompt produces a clear gain over standard prompting.}
  \label{tab:theory-framing}
  \small
  \setlength{\tabcolsep}{4pt}
  \begin{tabular}{lrrrr}
    \toprule
    \makecell[l]{Prompt\\type} & Standard & \makecell{Privacy\\Calculus} & \makecell{Bounded\\Rationality} & PMT \\
    \midrule
    \multicolumn{5}{l}{\emph{Gemini 3 Flash}} \\
    \texttt{none}  & 37.07 & 36.20 & 35.61 & 37.58 \\
    $E_u$    & 39.32 & 39.20 & 38.05 & 39.23 \\
    $A_u$    & 38.60 & 38.81 & 39.49 & 39.08 \\
    $E_u{+}A_u$ & 40.27 & 40.28 & 39.95 & 40.08 \\
    \midrule
    \multicolumn{5}{l}{\emph{Gemma 4 31B}} \\
    \texttt{none}  & 35.85 & 35.66 & 34.20 & 35.87 \\
    $E_u$    & 37.81 & 35.90 & 35.45 & 37.03 \\
    $A_u$    & 38.04 & 38.09 & 37.98 & 37.34 \\
    $E_u{+}A_u$ & 39.04 & 37.96 & 37.67 & 38.31 \\
    \bottomrule
  \end{tabular}
\end{wraptable}

\textbf{Privacy-informed prompts do not improve simulation quality.} We further ask whether conditioning LLMs to reason through well-established privacy theories before making privacy decisions would help improve simulation performance. We construct three privacy-informed variants of the persona prompt by inserting a \emph{privacy decision lens} block while leaving the rest of the prompt untouched. Each privacy decision lens is based on one of Privacy Calculus, Bounded Rationality, or Protection Motivation Theory. Full prompts can be found in Appendix~\ref{appendix:theory_prompts}.

Table~\ref{tab:theory-framing} reports accuracy on Gemini 3 Flash and Gemma 4 31B across the prompt types for all 5 studies. Across both models and every prompt type, no theory variant yields a meaningful gain. These results suggest that conditioning LLMs to reason about user privacy decisions through the lens of well-established privacy theories does not improve their ability to simulate individual privacy decisions beyond the signal already provided by the persona facets.
\section{Discussion and Conclusion}
\label{sec:discussion}
Conditioning off-the-shelf frontier LLMs to simulate users offers a promising way to obtain privacy judgments quickly and inexpensively, without running costly user studies. Consequently, a growing number of works are relying on LLM simulation/judgments as proxies for human privacy preferences and norms to be further used for evaluations. However, the results of our work suggest that these proxies may not be truly reflective of real users. 

We found that a user's demographics ($D_u$), previous experiences with AI ($E_u$), and general privacy attitudes ($A_u$) are a good starting point to provide modest simulation quality, but they are not enough to reliably align with users' mental models for privacy decisions~\citep{kang2015my}. For example, we would expect that a user's stated privacy attitudes would be the strongest predictor of privacy behavior, but our results show that this is not always the case. Moreover, low-stance / high-exposure users see only a $+1.07\%$ improvement from privacy attitudes and previous experiences and remain the hardest group to simulate. Our results show this is not a scaling issue, as varying model size and reasoning effort yields negligible gains, and prompting with well-established privacy theories does not help either. Together, these results indicate that the gap between what users \emph{say} about their privacy and what they actually \emph{do} is a fundamental challenge of simulating privacy behavior. We believe closing this gap requires more investigation into identifying richer and more nuanced user attributes, such as personality traits (Big Five)~\citep{john1999big}, as well as the user's rationale behind their privacy decision-making. Until then, we believe that LLM-simulated privacy behavior by itself cannot fully replace human judgments and should be complemented with human user studies.

A limitation of this work is that we tested only three persona facets: demographics, previous experiences, and privacy attitudes. This limitation is due to the small number of openly released user studies, and to the fact that these studies were not originally intended for user privacy simulation. Despite this limitation, \method{} provides a foundation for identifying what additional attributes might be necessary to bridge the gap between LLM simulation and real-world user behavior. 
\section*{Acknowledgments}
This work is supported by NSF award number DGE-1842487, NSF award number 2613950, and the REAL@USC-Meta center. We also thank the National Artificial Intelligence Research Resource (NAIRR) Pilot and CloudBank for contributing computing resources to this research.


\bibliographystyle{plainnat}
\bibliography{references}

\appendix

\etocdepthtag.toc{appendix}

\newpage
\etocsetnexttocdepth{subsubsection}
\etocsettagdepth{mainmatter}{none}
\etocsettagdepth{appendix}{subsubsection}
\renewcommand{\contentsname}{Appendix Contents}
\tableofcontents
\clearpage

\section{Persona Prompts}\label{appendix:persona_prompts}\label{app:prompts}

Every domain in \method{} uses a single prompt template with five labeled sections: a fixed \textsc{role} that frames the LLM as a participant simulator, a \textsc{user privacy persona} that injects the available facets ($D_u$, $E_u$, $A_u$) verbatim from the source study, a \textsc{task \& constraints} block that enforces JSON-only output, a \textsc{target output format} schema sized to the questionnaire, and the original \textsc{questionnaire} text reproduced from the source paper. Only the persona facets, the response scale, and the questionnaire body change across domains; the surrounding scaffolding is identical so that any cross-domain difference in simulation accuracy is attributable to the persona and questionnaire rather than to prompt engineering. Throughout this appendix, persona facets that the source study did not collect are rendered as \texttt{Not provided} (\eg demographics in \citet{liu2025prevalence} and \citet{tran2025understanding}). The five subsections below show one fully populated prompt for each domain, drawn from a real participant in the corresponding study.

\subsection{LLM Healthcare Consultation}
The prompt below corresponds to the LLM Healthcare Consultation domain~\citep{liu2025prevalence}.
\begin{promptbox}{Domain: LLM healthcare consultation \textbar{} Persona facets: $E_u$, $A_u$ \textbar{} Scale: 5-point Likert}
ROLE
You are an expert AI simulator acting as a specific human participant in a research study. Your task is to realistically answer a questionnaire about data-sharing scenarios exactly as this specific person would.

USER PRIVACY PERSONA
Internalize the user's traits and answer each scenario as this specific person would. Draw on whichever aspects of their profile are most relevant to the particular scenario.
- Domain Context: LLM healthcare consultation
- Demographics: Not provided.
- Previous AI Experiences: {'Which of the following best describes who you are when using the healthcare chatbot?': 'Users consulting on general health knowledge', 'Did you provide any personal information, such as age, in the consultation experience? Why?': 'No [Concerned about privacy; prefer not to disclose]', 'Do you have any healthcare consultation experience with any AI-large-scale-model except in our survey?': 'No', 'Do you have any experience with any AI-large-scale-model except in our survey?': 'Yes', 'Which AI-large-scale-model platforms have you used?': "ChatGPT | Bing Chat (Copilot) | Baidu's Ernie Bot | iFlytek's Spark | Alibaba's Qianwen", 'Have you ever heard of AI-large-scale-model?': '(Skipped)', 'Do you have any healthcare consultation experience online?': 'No', 'Why do you choose to use LLM-based healthcare consultation services?': "Save time with quick responses | Reduce healthcare costs | Protecting privacy (don't have to tell anyone) | Gain medical knowledge or references", "Why don't you want to use LLM-based healthcare consultation services?": '(Skipped)', 'Have you experienced a health privacy breach?': 'No'}
- Privacy Attitudes: {'Is online healthcare consultation more likely to compromise privacy than asking a doctor face-to-face?': 'Somewhat disagree', 'You will use reliable LLM-based healthcare consultation services in the future': 'Somewhat agree', 'You understand health privacy': 'Neutral', 'You are concerned about health privacy': 'Neutral'}

TASK & CONSTRAINTS
1. Answer all data-sharing scenarios in the QUESTIONNAIRE below from the perspective of the User Persona.
2. Return ONLY valid, parseable JSON.
3. DO NOT wrap your response in markdown code blocks (e.g., do not use ```json ... ```).
4. DO NOT add any commentary, greetings, or explanations outside the JSON object.
5. Keep field names, nesting, and data types strictly aligned with the TARGET OUTPUT FORMAT.

TARGET OUTPUT FORMAT (STRUCTURE ONLY - DO NOT COPY VALUES)
{
  "Q1": "<answer>",
  "Q2": "<answer>",
  "Q3": "<answer>",
  "Q4": "<answer>",
  "Q5": "<answer>",
  "Q6": "<answer>",
  "Q7": "<answer>",
  "Q8": "<answer>",
  "Q9": "<answer>",
  "Q10": "<answer>"
}

QUESTIONNAIRE
Imagine you use the AI-large-scale-model for healthcare consultation, and the conversation mentions your mental health information (e.g, stress levels, emotional fluctuations, psychotherapy, experiences). How acceptable is it for the following recipients to access this information?
Your options are 'Completely unacceptable', 'Somewhat unacceptable', 'Neutral', 'Somewhat acceptable', and 'Completely acceptable'.
Q1. Your family members.
Q2. Healthcare providers.
Q3. Government agencies.
Q4. LLM service providers.

How acceptable is it for Government agencies to access your mental health information under the following conditions?
Your options are 'Completely unacceptable', 'Somewhat unacceptable', 'Neutral', 'Somewhat acceptable', and 'Completely acceptable'.
Q5. If you have given consent,
Q6. If you are notified before collection,
Q7. If information is kept confidential and not stored,
Q8. If information will be stored,
Q9. If information is used for health monitoring,
Q10. If information is shared with others,
\end{promptbox}

\subsection{AI Agent Permissions}
The prompt below corresponds to the AI Agent Permissions domain~\citep{wu2025towards}.
\begin{promptbox}{Domain: AI agent permissions \textbar{} Persona facets: $D_u$, $E_u$, $A_u$ \textbar{} Scale: 4-point preference}
ROLE
You are an expert AI simulator acting as a specific human participant in a research study. Your task is to realistically answer a questionnaire about data-sharing scenarios exactly as this specific person would.

USER PRIVACY PERSONA
Internalize the user's traits and answer each scenario as this specific person would. Draw on whichever aspects of their profile are most relevant to the particular scenario.
- Domain Context: AI agent permissions
- Demographics: {'education': "Master's", 'age': '18-24', 'gender': 'Female'}
- Previous AI Experiences: {'On a scale from 1 (Not at all familiar) to 5 (Very familiar), how familiar are you with existing AI tools, such as ChatGPT and Gemini, in terms of your knowledge and usage? [1. Not at all familiar, 2. Slightly familiar, 3. Somewhat familiar, 4. Moderately familiar, 5. Very familiar]': '3', 'How often do you use these tools? [1. Never, 2. Rarely (e.g., once a month or less), 3. Occasionally (e.g., 1-2 times a week), 4. Frequently (e.g., several times a week), 5. Daily]': '2', 'For what purposes do you use these tools?': 'Mostly during school for help with work/generating ideas'}
- Privacy Attitudes: {'On a scale from 1 (Not at all) to 5 (Full trust), how much do you trust these tools to not violate your privacy?': '3', 'On a scale from 1 (Not important) to 5 (Extremely Important), how important is privacy to you when using digital tools or services, including AI tools?': '4', 'If you have any privacy concerns while using these tools, what are they?': 'The database being able to access personal data.', 'Select 3 domains you care the most about your privacy when using AI tools:': ['Social', 'Finance', 'Smart Home']}

TASK & CONSTRAINTS
1. Answer all data-sharing scenarios in the QUESTIONNAIRE below from the perspective of the User Persona.
2. Return ONLY valid, parseable JSON.
3. DO NOT wrap your response in markdown code blocks (e.g., do not use ```json ... ```).
4. DO NOT add any commentary, greetings, or explanations outside the JSON object.
5. Keep field names, nesting, and data types strictly aligned with the TARGET OUTPUT FORMAT.

TARGET OUTPUT FORMAT (STRUCTURE ONLY - DO NOT COPY VALUES)
{
  "Q1": "<answer>",
  "Q2": "<answer>",
  "Q3": "<answer>",
  "Q4": "<answer>",
  "Q5": "<answer>",
  "Q6": "<answer>",
  "Q7": "<answer>",
  "Q8": "<answer>",
  "Q9": "<answer>",
  "Q10": "<answer>"
}

QUESTIONNAIRE
In year 2049, AI personal assistants are an essential part of people's daily lives. These AI assistants handle everything - from helping with work, to managing finances, to organizing travel, and meal planning.
-> "Create a training plan for my upcoming marathon"
	- The AI assistant might consider user's health metrics and calendar to create a training plan
-> "Plan a stopover in Spain for a two-day vacation with my family on my way back to travel from Japan"
	- The AI assistant might consider user's calendar, their interests, and family info to suggest a plan
-> "Can you compose a bedtime story that my children will like?"
	- The AI assistant might consider children's interests to create a story
-> "Recommend a TV show or movie I'll love"
	- The AI assistant might consider user's viewing history and their mood for recommendations
Now imagine that you're living in year 2049! Your job is to train a personal assistant named "MIRA" for yourself. You will be presented several scenarios to personalize Mira.
Training MIRA to learn your data sharing preferences
Mira learns your data sharing preferences and automatically manages permissions on your behalf, so you never have to decide each one individually! Please go through the following scenarios and express your preferences! Let's begin with the queries:


USER: Plan my weekend in San Francisco. Please include surfing in the plan.
MIRA: Absolutely USER! I need the data listed below to address your query.
Please express your data sharing and accessing preferences. The permission preferences options are "Yes, always share", "Yes, but ask me next time", "No, but ask me next time", and "No, never share". Be careful, Mira or apps may collect incorrect or unnecessary data!

Q1. Data: Trip dates; App/MIRA: Weather
Q2. Data: Trip location; App/MIRA: Weather
Q3. Data: Trip dates; App/MIRA: Local Search
Q4. Data: Trip location; App/MIRA: Local Search
Q5. Data: Trip activities; App/MIRA: Local Search
Q6. Data: Email content; App/MIRA: Local Search

USER: Schedule a one-time payment of $1,200 to my landlord next Monday. Please label it "Rent Payment" and include my address for reference.
MIRA: Absolutely USER! I need the data listed below to address your query.
Please express your data sharing and accessing preferences. The permission preferences options are "Yes, always share", "Yes, but ask me next time", "No, but ask me next time", and "No, never share". Be careful, Mira or apps may collect incorrect or unnecessary data!

Q7. Data: Payment note; App/MIRA: Banking
Q8. Data: Account number; App/MIRA: Banking
Q9. Data: Account email; App/MIRA: Banking
Q10. Data: Account password; App/MIRA: Banking
\end{promptbox}

\subsection{LLM Chatbot}
The prompt below corresponds to the LLM Chatbot domain~\citep{tran2025understanding}.
\begin{promptbox}{Domain: LLM chatbot \textbar{} Persona facets: $E_u$, $A_u$ \textbar{} Scale: 0-100 numeric}
ROLE
You are an expert AI simulator acting as a specific human participant in a research study. Your task is to realistically answer a questionnaire about data-sharing scenarios exactly as this specific person would.

USER PRIVACY PERSONA
Internalize the user's traits and answer each scenario as this specific person would. Draw on whichever aspects of their profile are most relevant to the particular scenario.
- Domain Context: LLM chatbot
- Demographics: Not provided.
- Previous AI Experiences: {'How frequently do you use ChatGPT?': 'Daily', 'What is the date of your first use of ChatGPT?': '2023 or earlier', 'What is your ChatGPT subscription type?': 'ChatGPT Plus', 'Tasks ChatGPT is Used For': 'Daily Life Skills,Personal Education (e.g., learning new concepts),Searching for Information,Daily Life Skills (e.g., cooking and managing finances),Asking for Advice,Other', 'If you selected "Other" for tasks, please elaborate.': "Oddly, everything but writing which is what everyone else seems to use it for. I spend many hours asking about an attenborough doc Im watching etc. It's too helpful now so I pay for plus yea", 'Topics Discussed with ChatGPT': 'Health or Wellness Advice,Personal Finances,Mental Health Support,Travel, Hobbies,Entertainment (movies, videogames, etc.),Work/Job support,Scientific or academic information,Other ', 'If you selected "Other" for topics, please elaborate.': 'Chat GPT even helped me design and build a soundproof enclosure for my 3D printer and! a sound proof ventilation system for it which I was really happy with. I use it for tons now.'}
- Privacy Attitudes: {'Online companies collect too much personal information.  Please rate your agreement on a scale of 0 to 100 (0 = Strongly Disagree, 50 = Neutral, 100 = Strongly Agree).': 82, 'In general, I trust websites.  Please rate your agreement on a scale of 0 to 100 (0 = Strongly Disagree, 50 = Neutral, 100 = Strongly Agree).': 0, 'In general, I believe privacy is important.  Please rate your agreement on a scale of 0 to 100 (0 = Strongly Disagree, 50 = Neutral, 100 = Strongly Agree).': 100, 'I have privacy concerns about conversations with ChatGPT.  Please rate your agreement on a scale of 0 to 100 (0 = Strongly Disagree, 50 = Neutral, 100 = Strongly Agree).': 19, 'In general, I believe that my conversations with chatbots like ChatGPT will remain private.  Please rate your agreement on a scale of 0 to 100 (0 = Strongly Disagree, 50 = Neutral, 100 = Strongly Agree).': 40}

TASK & CONSTRAINTS
1. Answer all data-sharing scenarios in the QUESTIONNAIRE below from the perspective of the User Persona.
2. Return ONLY valid, parseable JSON.
3. DO NOT wrap your response in markdown code blocks (e.g., do not use ```json ... ```).
4. DO NOT add any commentary, greetings, or explanations outside the JSON object.
5. Keep field names, nesting, and data types strictly aligned with the TARGET OUTPUT FORMAT.

TARGET OUTPUT FORMAT (STRUCTURE ONLY - DO NOT COPY VALUES)
{
  "Q1": "<integer>",
  "Q2": "<integer>",
  "Q3": "<integer>",
  "Q4": "<integer>",
  "Q5": "<integer>",
  "Q6": "<integer>",
  "Q7": "<integer>",
  "Q8": "<integer>",
  "Q9": "<integer>",
  "Q10": "<integer>"
}

QUESTIONNAIRE
You will be shown a series of scenarios describing the use of your ChatGPT history. For each scenario, please indicate your agreement with the following statement: 'This is an appropriate use of my ChatGPT data.' Please rate your agreement on a scale of 0 to 100 (0 = Strongly Disagree, 50 = Neutral, 100 = Strongly Agree).
Q1. Your ChatGPT conversations were shared with a(n) a university computer science researcher based in China for the purpose of studying AI risks to society. Your conversations with ChatGPT about your personal health and wellness were shared. You were not informed that your data was collected. Your conversations were not anonymized before being shared. Personally identifying information from your conversations was not removed.
Q2. Your ChatGPT conversations were shared with a(n) a university social science researcher based in the US for the purpose of creating a public dataset for AI research. Your conversations with ChatGPT about social life and personal relationships were shared. You were not informed that your data was collected. Your conversations were not anonymized before being shared. Personally identifying information from your conversations was not removed.
Q3. Your ChatGPT conversations were shared with a(n) a charitable foundation based in the US for the purpose of creating a public dataset for AI research. All of your conversations with ChatGPT were shared. You were asked for consent in advance of the data being collected. Your conversations were anonymized before being shared. Personally identifying information from your conversations was automatically removed.
Q4. Your ChatGPT conversations were shared with a(n) a hospital based in the US for the purpose of personalizing advertising. Your conversations with ChatGPT where you used the model to help with your job were shared. You were not informed that your data was collected. Your conversations were not anonymized before being shared. Personally identifying information from your conversations was not removed.
Q5. Your ChatGPT conversations were shared with a(n) a university computer science researcher based in the European Union for the purpose of fighting terrorism. Your conversations with ChatGPT where you used the model to help with your job were shared. You were asked for consent in advance of the data being collected. Your conversations were not anonymized before being shared. Personally identifying information from your conversations was not removed.
Q6. Your ChatGPT conversations were shared with a(n) a university social science researcher based in the European Union for the purpose of training future AI models. Your conversations with ChatGPT where you sought legal or ethical guidance were shared. You were not informed that your data was collected. Your conversations were anonymized before being shared. Personally identifying information from your conversations was automatically removed.
Q7. Your ChatGPT conversations were shared with a(n) a university computer science researcher based in the European Union for the purpose of personalizing advertising. All of your conversations with ChatGPT were shared. You were not informed that your data was collected. Your conversations were anonymized before being shared. Personally identifying information from your conversations was automatically removed.
Q8. Your ChatGPT conversations were shared with a(n) a government agency based in the European Union for the purpose of predicting human behavior. Your conversations with ChatGPT about social life and personal relationships were shared. You were informed that your data was collected. Your conversations were not anonymized before being shared. Personally identifying information from your conversations was not removed.
Q9. Your ChatGPT conversations were shared with a(n) a big tech company based in the US for the purpose of studying AI risks to society. Your conversations with ChatGPT where you sought legal or ethical guidance were shared. You were not informed that your data was collected. Your conversations were anonymized before being shared. Personally identifying information from your conversations was automatically removed.
Q10. Your ChatGPT conversations were shared with a(n) a university social science researcher based in the European Union for the purpose of improving user experience with AI. Your conversations with ChatGPT where you used the model to help with your job were shared. You were informed that your data was collected. Your conversations were anonymized before being shared. Personally identifying information from your conversations was automatically removed.
\end{promptbox}

\subsection{Chatbot Group Chat}
The prompt below corresponds to the Chatbot Group Chat domain~\citep{chou2026bot}.
\begin{promptbox}{Domain: Chatbot group chat \textbar{} Persona facets: $D_u$, $E_u$, $A_u$ \textbar{} Scale: 5-point Likert}
ROLE
You are an expert AI simulator acting as a specific human participant in a research study. Your task is to realistically answer a questionnaire about data-sharing scenarios exactly as this specific person would.

USER PRIVACY PERSONA
Internalize the user's traits and answer each scenario as this specific person would. Draw on whichever aspects of their profile are most relevant to the particular scenario.
- Domain Context: Chatbot group chat
- Demographics: {"Age Group": "25 - 34", "Gender - Selected Choice": "Male", "Gender - Prefer to self-describe - Text": null, "Highest level of education completed": "Master's degree", "Technical work experience": "Yes", "Race and ethnicity - Selected Choice": "White or Caucasian", "Race and ethnicity - Prefer to self-describe - Text": null}
- Previous AI Experiences: {"Please select one instant messaging service from the following list that you use frequently (more than twice a week) and have seen chatbots on. If more than one applies, choose the one you use the most. The rest of the survey will be based on your selection here.": "Discord", "How often do you use the instant messaging service you selected in the previous question?": "Daily", "How often do you notice chatbots in group chats, including seeing them send messages or actively searching for their existence?": "Daily", "Please name 1-2 chatbots you encountered in group chats and describe them in 1-2 sentences.": "Cleverbot, a chatbot uses the inputs from users to learn and generate responses.\nXavier., a chatbot that can also interact with audios, images etc. It can also do internet searches for you.", "How often do you interact with chatbots in group chats, including actively tagging/mentioning chatbots in your messages, chatbots replying to your messages, or chatbots reading your messages?": "4-6 times per week", "What types of group chats have you seen chatbots in? Select all that apply. - Selected Choice": "Educational groups (classes, study groups),Friend or family groups,Public forums or community groups", "What types of group chats have you seen chatbots in? Select all that apply. - Other (please specify) - Text": null, "What purposes do you use chatbots for in group chats? Select all that apply.": "Productivity,Education,Entertainment & Games,Graphics & Design,Music,Social Networking", "Have you ever noticed any chatbots in private group chats you have participated in?": "Yes", "How often do you check for the presence of chatbots whenever you join private group chats?": "Rarely", "Regarding private group chats - Please elaborate on your answer to the previous question (how often you check for the presence of chatbots when joining private groups).": "I don't really go looking for the presence of chatbots, it's more something that happens by chance in the discord servers I join. I use them in the servers I'm already in.", "When do you check for the presence of chatbots in a private group? Select all that apply.": null, "Have you ever noticed any chatbots in public group chats you have participated in?": "Yes", "How often do you check for the presence of chatbots whenever you join public group chats?": "Rarely", "Regarding public group chats - Please elaborate on your answer to the previous question (how often you check for the presence of chatbots when joining public groups).": "I am not particularly interested in using chatbots in public chats that everyone can read, I rarely look for them there. I prefer the privacy of private bots, especially if the features are the same.", "When do you check for the presence of chatbots in a public group? Select all that apply.": null}
- Privacy Attitudes: {"To what extent are you familiar with what information chatbots on Discord can access in group chats?": "Very knowledgeable", "What type of messages do you believe a chatbot on Discord has access to?": "All messages sent in the chat, including those that are later deleted.", "Following up on the previous question, what information do you believe is included with the message a chatbot has access to on Discord? Select all that apply.": "The content of the message itself.,The message metadata, such as time sent.,The sender's username. A user has the same username in different groups.,The sender's pseudonym. A user has different pseudonyms in different groups.,The sender's metadata (other than the username), such as profile picture, gender, and email addresses.", "What type of information do you believe a chatbot on Discord can access about the group? Select all that apply.": "The name of the group.,The list of group members' usernames.,The profile pictures of group members.,The roles or statuses of group members (e.g., moderator)."}

TASK & CONSTRAINTS
1. Answer all data-sharing scenarios in the QUESTIONNAIRE below from the perspective of the User Persona.
2. Return ONLY valid, parseable JSON.
3. DO NOT wrap your response in markdown code blocks (e.g., do not use ```json ... ```).
4. DO NOT add any commentary, greetings, or explanations outside the JSON object.
5. Keep field names, nesting, and data types strictly aligned with the TARGET OUTPUT FORMAT.

TARGET OUTPUT FORMAT (STRUCTURE ONLY - DO NOT COPY VALUES)
{
  "Q1": "<answer>",
  "Q2": "<answer>",
  "Q3": "<answer>",
  "Q4": "<answer>",
  "Q5": "<answer>",
  "Q6": "<answer>",
  "Q7": "<answer>"
}

QUESTIONNAIRE
Please rate your level of agreement with the following statements about chatbots in group chats.
Your options are 'Strongly disagree', 'Somewhat disagree', 'Neither agree nor disagree', 'Somewhat agree', and 'Strongly agree'.

Q1. Chatbots from group chats with access to users' data as listed above would raise my privacy concerns.
Q2. I would be concerned that any personal information I share in the group chat could be collected by the chatbot and be used inappropriately.
Q3. I would stop sharing some sensitive information or talking about controversial topics in group chat because of privacy concerns about chatbots.
Q4. I would still use the chatbot after knowing what information it had access to.
Q5. I think chatbots accessing more information than necessary could cause privacy issues in public group chats.
Q6. I think chatbots accessing more information than necessary could cause privacy issues in private group chats.
Q7. I understand how chatbot providers protect the data collected from group chats.
\end{promptbox}

\subsection{LLM Conversational Agents}
The prompt below corresponds to the LLM Conversational Agents domain~\citep{zufferey2025ai}.
\begin{promptbox}{Domain: LLM conversational agents \textbar{} Persona facets: $D_u$, $E_u$, $A_u$ \textbar{} Scale: 5-point frequency}
ROLE
You are an expert AI simulator acting as a specific human participant in a research study. Your task is to realistically answer a questionnaire about data-sharing scenarios exactly as this specific person would.

USER PRIVACY PERSONA
Internalize the user's traits and answer each scenario as this specific person would. Draw on whichever aspects of their profile are most relevant to the particular scenario.
- Domain Context: LLM conversational agents
- Demographics: {'age': '18 - 29', 'country/region': 'Europe'}
- Previous AI Experiences: {'Are you using your own local chatbot instance based on only locally stored LLM model(s) (e.g., LLaMa, Mistral) and data?\n\n\n\nThis is about setting up your own chatbot. Apps such as the ChatGPT smartphone app are not included in this category.': 'Yes, but not exclusively (I also use chatbot services, as, for example, ChatGPT)', 'Among the following, for what purpose do you use your own local chatbot instance? - Selected Choice': 'General information seeking and learning,Entertainment (e.g., casual chats, text-based games)', 'Among the following, for what purpose do you use your own local chatbot instance? - Other (please specifiy) - Text': nan, 'Why do you have your own local chatbot instance and not only use chatbot services (e.g., ChatGPT, Copilot)? - Selected Choice': 'Performance (e.g., computation time, fine-tuning)', 'Why do you have your own local chatbot instance and not only use chatbot services (e.g., ChatGPT, Copilot)? - Other (please specifiy) - Text': nan, 'Please select the chatbots you usually use. - Selected Choice': 'ChatGPT', 'Please select the chatbots you usually use. - Other (please specify, use commas if multiple answers) - Text': nan, 'How frequently do you use chatbots?': 'A few times a week', 'On the days you are interacting with chatbots, how long does this interactions last on average (per day)?': 'between 1 and 2 hours', 'Among the following devices, on which one do you usually interact with chatbots? Several selections are possible. - Selected Choice': 'Laptop / Personal computer,Tablet,Smartphone', 'Among the following devices, on which one do you usually interact with chatbots? Several selections are possible. - Other (please specify) - Text': nan, 'Among the following, for what purpose do you usually use chatbots? Several selections are possible. - Selected Choice': 'General information seeking and learning,Programming support', 'Among the following, for what purpose do you usually use chatbots? Several selections are possible. - Other - Text': nan}
- Privacy Attitudes: {'Please indicate to what extent you agree with each of the following statements [Strongly agree, Disgree, Moderately disagree, Neither agree nor disagree, Moderately agree, Agree, Strongly Agree]. - It usually bothers me when online companies ask me for personal information.': 'Agree', 'Please indicate to what extent you agree with each of the following statements [Strongly agree, Disgree, Moderately disagree, Neither agree nor disagree, Moderately agree, Agree, Strongly Agree]. - When online companies ask me for personal information, I sometimes think twice before providing it.': 'Moderately agree', 'Please indicate to what extent you agree with each of the following statements [Strongly agree, Disgree, Moderately disagree, Neither agree nor disagree, Moderately agree, Agree, Strongly Agree]. - It bothers me to give personal information to so many online companies.': 'Agree', 'Please indicate to what extent you agree with each of the following statements [Strongly agree, Disgree, Moderately disagree, Neither agree nor disagree, Moderately agree, Agree, Strongly Agree]. - I am concerned that online companies are collecting too much personal information about me.': 'Agree', 'Please indicate to what extent you agree with each of the following statements [Strongly agree, Disgree, Moderately disagree, Neither agree nor disagree, Moderately agree, Agree, Strongly Agree]. - Consumer online privacy is really a matter of consumers' right to exercise control and autonomy over decisions about how their information is collected, used, and shared.': 'Moderately agree', 'Please indicate to what extent you agree with each of the following statements [Strongly agree, Disgree, Moderately disagree, Neither agree nor disagree, Moderately agree, Agree, Strongly Agree]. - Consumer control of personal information lies at the heart of consumer privacy.': 'Moderately agree', 'Please indicate to what extent you agree with each of the following statements [Strongly agree, Disgree, Moderately disagree, Neither agree nor disagree, Moderately agree, Agree, Strongly Agree]. - Companies seeking information online should disclose the way the data are collected, processed, and used.': 'Agree', 'Please indicate to what extent you agree with each of the following statements [Strongly agree, Disgree, Moderately disagree, Neither agree nor disagree, Moderately agree, Agree, Strongly Agree]. - A good consumer online privacy policy should have a clear and conspicuous disclosure.': 'Agree'}

TASK & CONSTRAINTS
1. Answer all data-sharing scenarios in the QUESTIONNAIRE below from the perspective of the User Persona.
2. Return ONLY valid, parseable JSON.
3. DO NOT wrap your response in markdown code blocks (e.g., do not use ```json ... ```).
4. DO NOT add any commentary, greetings, or explanations outside the JSON object.
5. Keep field names, nesting, and data types strictly aligned with the TARGET OUTPUT FORMAT.

TARGET OUTPUT FORMAT (STRUCTURE ONLY - DO NOT COPY VALUES)
{
  "Q1": "<answer>",
  "Q2": "<answer>",
  "Q3": "<answer>",
  "Q4": "<answer>",
  "Q5": "<answer>",
  "Q6": "<answer>",
  "Q7": "<answer>",
  "Q8": "<answer>",
  "Q9": "<answer>",
  "Q10": "<answer>"
}

QUESTIONNAIRE
In the following part of the survey, we will ask you question about how frequently you share different type of personal data with chatbots. To evaluate this frequency, please think about the number of times where it would have been relevant to share a given type of data. For example, you should select 'I usually share' if you usually share a given type of data when it is relevant for your chatbot usage at the moment. If you never shared a given type of data in any context, please just answer 'I never shared'
For all the following types of data related to identifiers and general information, please indicate how frequently you share such information with a chatbot.
Your options are 'I never shared', 'I rarely shared', 'I sometimes shared', 'I often shared', and 'I usually shared'.
Information about your...
Q1. Personal identifiers (e.g., name, ID card number, email address, phone number)
Q2. Account credentials (i.e., usernames, passwords)
Q3. Location and mobility (e.g., homeplace, workplace, current location)
Q4. Ethnicity and/or citizenship
Q5. Criminal records

For all the following types of data about standard of living and opinions, please indicate how frequently you share such information with a chatbot?
Your options are 'I never shared', 'I rarely shared', 'I sometimes shared', 'I often shared', and 'I usually shared'.
Information about your...
Q6. Family life and relationship (e.g., civil status, number of children)
Q7. Wealth details (e.g., salary, savings amount, land ownership)
Q8. Job / Education
Q9. Religion
Q10. Political views
\end{promptbox}

\subsection{Privacy-informed Prompt Variants for LLM Conversational Agents}\label{appendix:theory_prompts}

Below are the three privacy-informed prompts for the same participant from \citet{zufferey2025ai} as in Appendix~\ref{appendix:persona_prompts} above. Each privacy-informed prompt is based on the original prompt by adding a \textsc{privacy decision lens} block inserted between the persona section and the task instructions; the role, persona content, output format, and questionnaire are identical across the three privacy-informed prompts and the original prompt.

\subsubsection{Privacy Calculus}
\begin{promptbox}{Theory: Privacy Calculus \textbar{} Domain: LLM conversational agents \textbar{} Persona facets: $E_u$, $A_u$ \textbar{} Scale: 5-point frequency}
ROLE
You are an expert AI simulator acting as a specific human participant in a research study. Your task is to realistically answer a questionnaire about data-sharing scenarios exactly as this specific person would.

USER PRIVACY PERSONA
Internalize the user's traits and answer each scenario as this specific person would. Draw on whichever aspects of their profile are most relevant to the particular scenario.
- Domain Context: LLM conversational agents
- Demographics: Not provided.
- Previous AI Experiences: {'Are you using your own local chatbot instance based on only locally stored LLM model(s) (e.g., LLaMa, Mistral) and data?\n\n\n\nThis is about setting up your own chatbot. Apps such as the ChatGPT smartphone app are not included in this category.': 'Yes, but not exclusively (I also use chatbot services, as, for example, ChatGPT)', 'Among the following, for what purpose do you use your own local chatbot instance? - Selected Choice': 'General information seeking and learning,Entertainment (e.g., casual chats, text-based games)', 'Among the following, for what purpose do you use your own local chatbot instance? - Other (please specifiy) - Text': nan, 'Why do you have your own local chatbot instance and not only use chatbot services (e.g., ChatGPT, Copilot)? - Selected Choice': 'Performance (e.g., computation time, fine-tuning)', 'Why do you have your own local chatbot instance and not only use chatbot services (e.g., ChatGPT, Copilot)? - Other (please specifiy) - Text': nan, 'Please select the chatbots you usually use. - Selected Choice': 'ChatGPT', 'Please select the chatbots you usually use. - Other (please specify, use commas if multiple answers) - Text': nan, 'How frequently do you use chatbots?': 'A few times a week', 'On the days you are interacting with chatbots, how long does this interactions last on average (per day)?': 'between 1 and 2 hours', 'Among the following devices, on which one do you usually interact with chatbots? Several selections are possible. - Selected Choice': 'Laptop / Personal computer,Tablet,Smartphone', 'Among the following devices, on which one do you usually interact with chatbots? Several selections are possible. - Other (please specify) - Text': nan, 'Among the following, for what purpose do you usually use chatbots? Several selections are possible. - Selected Choice': 'General information seeking and learning,Programming support', 'Among the following, for what purpose do you usually use chatbots? Several selections are possible. - Other - Text': nan}
- Privacy Attitudes: {'Please indicate to what extent you agree with each of the following statements [Strongly agree, Disgree, Moderately disagree, Neither agree nor disagree, Moderately agree, Agree, Strongly Agree]. - It usually bothers me when online companies ask me for personal information.': 'Agree', 'Please indicate to what extent you agree with each of the following statements [Strongly agree, Disgree, Moderately disagree, Neither agree nor disagree, Moderately agree, Agree, Strongly Agree]. - When online companies ask me for personal information, I sometimes think twice before providing it.': 'Moderately agree', 'Please indicate to what extent you agree with each of the following statements [Strongly agree, Disgree, Moderately disagree, Neither agree nor disagree, Moderately agree, Agree, Strongly Agree]. - It bothers me to give personal information to so many online companies.': 'Agree', 'Please indicate to what extent you agree with each of the following statements [Strongly agree, Disgree, Moderately disagree, Neither agree nor disagree, Moderately agree, Agree, Strongly Agree]. - I am concerned that online companies are collecting too much personal information about me.': 'Agree', 'Please indicate to what extent you agree with each of the following statements [Strongly agree, Disgree, Moderately disagree, Neither agree nor disagree, Moderately agree, Agree, Strongly Agree]. - Consumer online privacy is really a matter of consumers' right to exercise control and autonomy over decisions about how their information is collected, used, and shared.': 'Moderately agree', 'Please indicate to what extent you agree with each of the following statements [Strongly agree, Disgree, Moderately disagree, Neither agree nor disagree, Moderately agree, Agree, Strongly Agree]. - Consumer control of personal information lies at the heart of consumer privacy.': 'Moderately agree', 'Please indicate to what extent you agree with each of the following statements [Strongly agree, Disgree, Moderately disagree, Neither agree nor disagree, Moderately agree, Agree, Strongly Agree]. - Companies seeking information online should disclose the way the data are collected, processed, and used.': 'Agree', 'Please indicate to what extent you agree with each of the following statements [Strongly agree, Disgree, Moderately disagree, Neither agree nor disagree, Moderately agree, Agree, Strongly Agree]. - A good consumer online privacy policy should have a clear and conspicuous disclosure.': 'Agree'}

PRIVACY DECISION LENS -- PRIVACY CALCULUS
Evaluate each scenario as this person performing a deliberate cost-benefit calculation about disclosure. For each request:
1. Identify the perceived BENEFITS of sharing -- service utility, personalization quality, convenience, time saved, social or relational value.
2. Identify the perceived RISKS or COSTS of sharing -- loss of control over the data, downstream reuse or profiling, embarrassment, identifiability, future harm, regulatory or employer exposure.
3. Weigh the two through the lens of this person's stated attitudes and prior experiences; their answer should reflect whichever side they judge to dominate.
4. When benefits clearly outweigh perceived risks for them, lean toward acceptance; when risks dominate, lean toward refusal; when the two are comparable, lean toward a conditional or intermediate option.
Pick the answer this person would arrive at after this tradeoff, not the answer a privacy-maximizing or utility-maximizing agent would give.

TASK & CONSTRAINTS
1. Answer all data-sharing scenarios in the QUESTIONNAIRE below from the perspective of the User Persona.
2. Return ONLY valid, parseable JSON.
3. DO NOT wrap your response in markdown code blocks (e.g., do not use ```json ... ```).
4. DO NOT add any commentary, greetings, or explanations outside the JSON object.
5. Keep field names, nesting, and data types strictly aligned with the TARGET OUTPUT FORMAT.

TARGET OUTPUT FORMAT (STRUCTURE ONLY - DO NOT COPY VALUES)
{
  "Q1": "<answer>",
  "Q2": "<answer>",
  "Q3": "<answer>",
  "Q4": "<answer>",
  "Q5": "<answer>",
  "Q6": "<answer>",
  "Q7": "<answer>",
  "Q8": "<answer>",
  "Q9": "<answer>",
  "Q10": "<answer>"
}

QUESTIONNAIRE
In the following part of the survey, we will ask you question about how frequently you share different type of personal data with chatbots. To evaluate this frequency, please think about the number of times where it would have been relevant to share a given type of data. For example, you should select 'I usually share' if you usually share a given type of data when it is relevant for your chatbot usage at the moment. If you never shared a given type of data in any context, please just answer 'I never shared'
For all the following types of data related to identifiers and general information, please indicate how frequently you share such information with a chatbot.
Your options are 'I never shared', 'I rarely shared', 'I sometimes shared', 'I often shared', and 'I usually shared'.
Information about your...
Q1. Personal identifiers (e.g., name, ID card number, email address, phone number)
Q2. Account credentials (i.e., usernames, passwords)
Q3. Location and mobility (e.g., homeplace, workplace, current location)
Q4. Ethnicity and/or citizenship
Q5. Criminal records

For all the following types of data about standard of living and opinions, please indicate how frequently you share such information with a chatbot?
Your options are 'I never shared', 'I rarely shared', 'I sometimes shared', 'I often shared', and 'I usually shared'.
Information about your...
Q6. Family life and relationship (e.g., civil status, number of children)
Q7. Wealth details (e.g., salary, savings amount, land ownership)
Q8. Job / Education
Q9. Religion
Q10. Political views
\end{promptbox}

\subsubsection{Bounded Rationality}
\begin{promptbox}{Theory: Bounded Rationality \textbar{} Domain: LLM conversational agents \textbar{} Persona facets: $E_u$, $A_u$ \textbar{} Scale: 5-point frequency}
ROLE
You are an expert AI simulator acting as a specific human participant in a research study. Your task is to realistically answer a questionnaire about data-sharing scenarios exactly as this specific person would.

USER PRIVACY PERSONA
Internalize the user's traits and answer each scenario as this specific person would. Draw on whichever aspects of their profile are most relevant to the particular scenario.
- Domain Context: LLM conversational agents
- Demographics: Not provided.
- Previous AI Experiences: {'Are you using your own local chatbot instance based on only locally stored LLM model(s) (e.g., LLaMa, Mistral) and data?\n\n\n\nThis is about setting up your own chatbot. Apps such as the ChatGPT smartphone app are not included in this category.': 'Yes, but not exclusively (I also use chatbot services, as, for example, ChatGPT)', 'Among the following, for what purpose do you use your own local chatbot instance? - Selected Choice': 'General information seeking and learning,Entertainment (e.g., casual chats, text-based games)', 'Among the following, for what purpose do you use your own local chatbot instance? - Other (please specifiy) - Text': nan, 'Why do you have your own local chatbot instance and not only use chatbot services (e.g., ChatGPT, Copilot)? - Selected Choice': 'Performance (e.g., computation time, fine-tuning)', 'Why do you have your own local chatbot instance and not only use chatbot services (e.g., ChatGPT, Copilot)? - Other (please specifiy) - Text': nan, 'Please select the chatbots you usually use. - Selected Choice': 'ChatGPT', 'Please select the chatbots you usually use. - Other (please specify, use commas if multiple answers) - Text': nan, 'How frequently do you use chatbots?': 'A few times a week', 'On the days you are interacting with chatbots, how long does this interactions last on average (per day)?': 'between 1 and 2 hours', 'Among the following devices, on which one do you usually interact with chatbots? Several selections are possible. - Selected Choice': 'Laptop / Personal computer,Tablet,Smartphone', 'Among the following devices, on which one do you usually interact with chatbots? Several selections are possible. - Other (please specify) - Text': nan, 'Among the following, for what purpose do you usually use chatbots? Several selections are possible. - Selected Choice': 'General information seeking and learning,Programming support', 'Among the following, for what purpose do you usually use chatbots? Several selections are possible. - Other - Text': nan}
- Privacy Attitudes: {'Please indicate to what extent you agree with each of the following statements [Strongly agree, Disgree, Moderately disagree, Neither agree nor disagree, Moderately agree, Agree, Strongly Agree]. - It usually bothers me when online companies ask me for personal information.': 'Agree', 'Please indicate to what extent you agree with each of the following statements [Strongly agree, Disgree, Moderately disagree, Neither agree nor disagree, Moderately agree, Agree, Strongly Agree]. - When online companies ask me for personal information, I sometimes think twice before providing it.': 'Moderately agree', 'Please indicate to what extent you agree with each of the following statements [Strongly agree, Disgree, Moderately disagree, Neither agree nor disagree, Moderately agree, Agree, Strongly Agree]. - It bothers me to give personal information to so many online companies.': 'Agree', 'Please indicate to what extent you agree with each of the following statements [Strongly agree, Disgree, Moderately disagree, Neither agree nor disagree, Moderately agree, Agree, Strongly Agree]. - I am concerned that online companies are collecting too much personal information about me.': 'Agree', 'Please indicate to what extent you agree with each of the following statements [Strongly agree, Disgree, Moderately disagree, Neither agree nor disagree, Moderately agree, Agree, Strongly Agree]. - Consumer online privacy is really a matter of consumers' right to exercise control and autonomy over decisions about how their information is collected, used, and shared.': 'Moderately agree', 'Please indicate to what extent you agree with each of the following statements [Strongly agree, Disgree, Moderately disagree, Neither agree nor disagree, Moderately agree, Agree, Strongly Agree]. - Consumer control of personal information lies at the heart of consumer privacy.': 'Moderately agree', 'Please indicate to what extent you agree with each of the following statements [Strongly agree, Disgree, Moderately disagree, Neither agree nor disagree, Moderately agree, Agree, Strongly Agree]. - Companies seeking information online should disclose the way the data are collected, processed, and used.': 'Agree', 'Please indicate to what extent you agree with each of the following statements [Strongly agree, Disgree, Moderately disagree, Neither agree nor disagree, Moderately agree, Agree, Strongly Agree]. - A good consumer online privacy policy should have a clear and conspicuous disclosure.': 'Agree'}

PRIVACY DECISION LENS -- BOUNDED RATIONALITY
Evaluate each scenario as this person under realistic cognitive constraints, not as a perfectly rational analyst:
- People rarely enumerate every risk. They rely on mental shortcuts: "does this look like the kind of thing I usually allow?", "is this the default?", "the system asked, so it probably needs it", "everyone seems to use this".
- They satisfice -- accept the first option that feels good enough -- rather than optimize.
- Stated privacy attitudes and concrete behavior can diverge (the privacy paradox). This person's answer in a specific scenario may not perfectly match their abstract stance, particularly when the request feels routine, low-friction, or socially normal.
- If the request feels familiar given their previous AI experiences, the heuristic favors acceptance even when their attitudes are cautious. If the request feels unfamiliar, intrusive, or salient, the heuristic flips toward refusal.
Pick the option this person -- as a cognitively bounded, heuristic-driven decision-maker -- would most plausibly land on in the moment, not the option a fully deliberative version of them would compute.

TASK & CONSTRAINTS
1. Answer all data-sharing scenarios in the QUESTIONNAIRE below from the perspective of the User Persona.
2. Return ONLY valid, parseable JSON.
3. DO NOT wrap your response in markdown code blocks (e.g., do not use ```json ... ```).
4. DO NOT add any commentary, greetings, or explanations outside the JSON object.
5. Keep field names, nesting, and data types strictly aligned with the TARGET OUTPUT FORMAT.

TARGET OUTPUT FORMAT (STRUCTURE ONLY - DO NOT COPY VALUES)
{
  "Q1": "<answer>",
  "Q2": "<answer>",
  "Q3": "<answer>",
  "Q4": "<answer>",
  "Q5": "<answer>",
  "Q6": "<answer>",
  "Q7": "<answer>",
  "Q8": "<answer>",
  "Q9": "<answer>",
  "Q10": "<answer>"
}

QUESTIONNAIRE
In the following part of the survey, we will ask you question about how frequently you share different type of personal data with chatbots. To evaluate this frequency, please think about the number of times where it would have been relevant to share a given type of data. For example, you should select 'I usually share' if you usually share a given type of data when it is relevant for your chatbot usage at the moment. If you never shared a given type of data in any context, please just answer 'I never shared'
For all the following types of data related to identifiers and general information, please indicate how frequently you share such information with a chatbot.
Your options are 'I never shared', 'I rarely shared', 'I sometimes shared', 'I often shared', and 'I usually shared'.
Information about your...
Q1. Personal identifiers (e.g., name, ID card number, email address, phone number)
Q2. Account credentials (i.e., usernames, passwords)
Q3. Location and mobility (e.g., homeplace, workplace, current location)
Q4. Ethnicity and/or citizenship
Q5. Criminal records

For all the following types of data about standard of living and opinions, please indicate how frequently you share such information with a chatbot?
Your options are 'I never shared', 'I rarely shared', 'I sometimes shared', 'I often shared', and 'I usually shared'.
Information about your...
Q6. Family life and relationship (e.g., civil status, number of children)
Q7. Wealth details (e.g., salary, savings amount, land ownership)
Q8. Job / Education
Q9. Religion
Q10. Political views
\end{promptbox}

\subsubsection{Protection Motivation Theory}
\begin{promptbox}{Theory: Protection Motivation Theory \textbar{} Domain: LLM conversational agents \textbar{} Persona facets: $E_u$, $A_u$ \textbar{} Scale: 5-point frequency}
ROLE
You are an expert AI simulator acting as a specific human participant in a research study. Your task is to realistically answer a questionnaire about data-sharing scenarios exactly as this specific person would.

USER PRIVACY PERSONA
Internalize the user's traits and answer each scenario as this specific person would. Draw on whichever aspects of their profile are most relevant to the particular scenario.
- Domain Context: LLM conversational agents
- Demographics: Not provided.
- Previous AI Experiences: {'Are you using your own local chatbot instance based on only locally stored LLM model(s) (e.g., LLaMa, Mistral) and data?\n\n\n\nThis is about setting up your own chatbot. Apps such as the ChatGPT smartphone app are not included in this category.': 'Yes, but not exclusively (I also use chatbot services, as, for example, ChatGPT)', 'Among the following, for what purpose do you use your own local chatbot instance? - Selected Choice': 'General information seeking and learning,Entertainment (e.g., casual chats, text-based games)', 'Among the following, for what purpose do you use your own local chatbot instance? - Other (please specifiy) - Text': nan, 'Why do you have your own local chatbot instance and not only use chatbot services (e.g., ChatGPT, Copilot)? - Selected Choice': 'Performance (e.g., computation time, fine-tuning)', 'Why do you have your own local chatbot instance and not only use chatbot services (e.g., ChatGPT, Copilot)? - Other (please specifiy) - Text': nan, 'Please select the chatbots you usually use. - Selected Choice': 'ChatGPT', 'Please select the chatbots you usually use. - Other (please specify, use commas if multiple answers) - Text': nan, 'How frequently do you use chatbots?': 'A few times a week', 'On the days you are interacting with chatbots, how long does this interactions last on average (per day)?': 'between 1 and 2 hours', 'Among the following devices, on which one do you usually interact with chatbots? Several selections are possible. - Selected Choice': 'Laptop / Personal computer,Tablet,Smartphone', 'Among the following devices, on which one do you usually interact with chatbots? Several selections are possible. - Other (please specify) - Text': nan, 'Among the following, for what purpose do you usually use chatbots? Several selections are possible. - Selected Choice': 'General information seeking and learning,Programming support', 'Among the following, for what purpose do you usually use chatbots? Several selections are possible. - Other - Text': nan}
- Privacy Attitudes: {'Please indicate to what extent you agree with each of the following statements [Strongly agree, Disgree, Moderately disagree, Neither agree nor disagree, Moderately agree, Agree, Strongly Agree]. - It usually bothers me when online companies ask me for personal information.': 'Agree', 'Please indicate to what extent you agree with each of the following statements [Strongly agree, Disgree, Moderately disagree, Neither agree nor disagree, Moderately agree, Agree, Strongly Agree]. - When online companies ask me for personal information, I sometimes think twice before providing it.': 'Moderately agree', 'Please indicate to what extent you agree with each of the following statements [Strongly agree, Disgree, Moderately disagree, Neither agree nor disagree, Moderately agree, Agree, Strongly Agree]. - It bothers me to give personal information to so many online companies.': 'Agree', 'Please indicate to what extent you agree with each of the following statements [Strongly agree, Disgree, Moderately disagree, Neither agree nor disagree, Moderately agree, Agree, Strongly Agree]. - I am concerned that online companies are collecting too much personal information about me.': 'Agree', 'Please indicate to what extent you agree with each of the following statements [Strongly agree, Disgree, Moderately disagree, Neither agree nor disagree, Moderately agree, Agree, Strongly Agree]. - Consumer online privacy is really a matter of consumers' right to exercise control and autonomy over decisions about how their information is collected, used, and shared.': 'Moderately agree', 'Please indicate to what extent you agree with each of the following statements [Strongly agree, Disgree, Moderately disagree, Neither agree nor disagree, Moderately agree, Agree, Strongly Agree]. - Consumer control of personal information lies at the heart of consumer privacy.': 'Moderately agree', 'Please indicate to what extent you agree with each of the following statements [Strongly agree, Disgree, Moderately disagree, Neither agree nor disagree, Moderately agree, Agree, Strongly Agree]. - Companies seeking information online should disclose the way the data are collected, processed, and used.': 'Agree', 'Please indicate to what extent you agree with each of the following statements [Strongly agree, Disgree, Moderately disagree, Neither agree nor disagree, Moderately agree, Agree, Strongly Agree]. - A good consumer online privacy policy should have a clear and conspicuous disclosure.': 'Agree'}

PRIVACY DECISION LENS -- PROTECTION MOTIVATION THEORY
Evaluate each scenario as this person running two parallel appraisals:

THREAT APPRAISAL
- Perceived severity: how harmful would exposure of this specific data be for this person, given their profile (e.g., medical, financial, identity, employment-relevant)?
- Perceived vulnerability: how likely does this person believe a harmful outcome -- leak, misuse, profiling, secondary use -- actually is in this context?

COPING APPRAISAL
- Response efficacy: does refusing, redacting, or limiting disclosure actually reduce the threat, in this person's belief?
- Self-efficacy: does this person feel able to refuse or push back without losing the service or the relationship?
- Response cost: what do they give up -- utility, convenience, social standing, continuity of care -- by protecting?

Combine the appraisals. High threat x high coping efficacy -> protective behavior (refusal or restriction). Low threat, or high threat but low coping efficacy / high response cost -> likely acceptance, sometimes reluctantly. Intermediate cases -> conditional acceptance. Pick the answer this person -- given their stated attitudes, demographics, and previous experiences -- would arrive at after this combined appraisal.

TASK & CONSTRAINTS
1. Answer all data-sharing scenarios in the QUESTIONNAIRE below from the perspective of the User Persona.
2. Return ONLY valid, parseable JSON.
3. DO NOT wrap your response in markdown code blocks (e.g., do not use ```json ... ```).
4. DO NOT add any commentary, greetings, or explanations outside the JSON object.
5. Keep field names, nesting, and data types strictly aligned with the TARGET OUTPUT FORMAT.

TARGET OUTPUT FORMAT (STRUCTURE ONLY - DO NOT COPY VALUES)
{
  "Q1": "<answer>",
  "Q2": "<answer>",
  "Q3": "<answer>",
  "Q4": "<answer>",
  "Q5": "<answer>",
  "Q6": "<answer>",
  "Q7": "<answer>",
  "Q8": "<answer>",
  "Q9": "<answer>",
  "Q10": "<answer>"
}

QUESTIONNAIRE
In the following part of the survey, we will ask you question about how frequently you share different type of personal data with chatbots. To evaluate this frequency, please think about the number of times where it would have been relevant to share a given type of data. For example, you should select 'I usually share' if you usually share a given type of data when it is relevant for your chatbot usage at the moment. If you never shared a given type of data in any context, please just answer 'I never shared'
For all the following types of data related to identifiers and general information, please indicate how frequently you share such information with a chatbot.
Your options are 'I never shared', 'I rarely shared', 'I sometimes shared', 'I often shared', and 'I usually shared'.
Information about your...
Q1. Personal identifiers (e.g., name, ID card number, email address, phone number)
Q2. Account credentials (i.e., usernames, passwords)
Q3. Location and mobility (e.g., homeplace, workplace, current location)
Q4. Ethnicity and/or citizenship
Q5. Criminal records

For all the following types of data about standard of living and opinions, please indicate how frequently you share such information with a chatbot?
Your options are 'I never shared', 'I rarely shared', 'I sometimes shared', 'I often shared', and 'I usually shared'.
Information about your...
Q6. Family life and relationship (e.g., civil status, number of children)
Q7. Wealth details (e.g., salary, savings amount, land ownership)
Q8. Job / Education
Q9. Religion
Q10. Political views
\end{promptbox}

\section{Example No-Persona Reasoning Trace}\label{appendix:reasoning_trace}
We show one reasoning trace (summarized because OpenAI does not provide full traces) from GPT-5.4 (medium reasoning effort) on the LLM Healthcare Consultation domain at the no-persona prompt type (\texttt{none}). The trace is taken verbatim from the OpenAI Responses API \texttt{summary\_text} blocks. Note how, absent any user-specific signal, the model explicitly chooses to ``simulate a generic person with moderate privacy attitudes'' and reasons in terms of population-level privacy norms (consent, confidentiality, recipient sensitivity) rather than properties of any individual. 

\begin{promptbox}{Reasoning trace: GPT-5.4 (medium reasoning) \textbar{} Domain: LLM Healthcare Consultation \textbar{} Prompt type: \texttt{none} (no persona)}
**Considering privacy in healthcare**

I need to provide an answer as a persona related to healthcare consultation, but I have limited information. It seems I should simulate a generic person with moderate privacy attitudes, especially in mental health. I'll focus on constructing the answer using JSON format and aim for values that are consistent with typical privacy concerns. It appears that consent and confidentiality are crucial, but sharing information with family or authorities isn't acceptable. I should maintain this perspective throughout.

**Creating a cautious privacy stance**

I want to craft a balanced and cautious stance regarding privacy in healthcare. For family members, I find sharing is somewhat unacceptable, while sharing with healthcare providers is completely acceptable. Government agencies are entirely unacceptable. LLM service providers also fall into the somewhat unacceptable category. With consent, sharing becomes somewhat acceptable, and being notified before data collection is neutral. Keeping data confidential yet not stored is somewhat acceptable, but storing is somewhat unacceptable. Sharing information with others is completely unacceptable. I need to format this in valid JSON with strings.
\end{promptbox}

\begin{promptbox}{Final answer (10-question 5-pt acceptability schema, $r_u$ omitted)}
{
  "Q1": "Somewhat unacceptable",
  "Q2": "Completely acceptable",
  "Q3": "Completely unacceptable",
  "Q4": "Somewhat unacceptable",
  "Q5": "Somewhat acceptable",
  "Q6": "Neutral",
  "Q7": "Somewhat acceptable",
  "Q8": "Somewhat unacceptable",
  "Q9": "Somewhat acceptable",
  "Q10": "Completely unacceptable"
}
\end{promptbox}

\section{More Details About User Studies}\label{appendix:user_studies}

This appendix gives a per-study description of the five user studies we draw from, summarizing the data-sharing context, the participant pool, what was collected for each persona facet, and the original key findings.

\subsection{LLM Healthcare Consultation}
\paragraph{Context.} \citet{liu2025prevalence} elicit privacy norms for sharing mental-health information mentioned in an LLM-based healthcare consultation. Each questionnaire asks how acceptable it is for four recipients (family members, healthcare providers, government agencies, LLM service providers) to access this information, and then probes acceptability of access by one specific recipient under six contextual conditions (consent, notification, confidentiality, storage, health-monitoring use, onward sharing) on a five-point Likert scale.
\paragraph{Participants.} 846 adult users from China recruited online. We restrict our analysis to participants who answered questions regarding mental health information or medical information because of greater variance in the responses. Hence, there are 417 participants. Demographics were not released in the public dataset, so $D_u = \emptyset$ for this domain.
\paragraph{Persona facets.} $E_u$ captures the participant's role when consulting (general health vs.\ personal issue vs.\ medical professional), prior LLM and online-consultation experience, motivations for/against using LLM healthcare services, and whether they have previously experienced a health-data breach. $A_u$ records four agreement items on health-privacy understanding, concern, and willingness to use LLM consultation.
\paragraph{Key findings.} Despite high stated privacy concern, the prevalence of LLM use overshadows those concerns: many participants disclose personal information to the chatbot anyway, motivated by convenience and a (sometimes mistaken) belief that the LLM will keep information confidential. Acceptability of disclosure to a recipient is strongly conditioned on consent and confidentiality, but drops sharply when the data will be stored or shared onward.
\paragraph{Data availability.} The raw user-study data is released by the original authors at \url{https://github.com/Cristliu/LLMHealthPrivacy_UserStudy} under the MIT License.
\begin{tcolorbox}[
    enhanced,
    breakable,
    colback=gray!3,
    colframe=black!55,
    boxrule=0.5pt,
    arc=2pt,
    left=6pt,right=6pt,top=4pt,bottom=4pt,
    fonttitle=\bfseries\small,
    coltitle=black,
    colbacktitle=gray!12,
    title={Example: User from LLM Healthcare Consultation \citep{liu2025prevalence}},
    fontupper=\small,
]
\textbf{Demographics ($D_u$).}\quad \texttt{Not provided.}\\[2pt]
\textbf{Previous experiences ($E_u$).}\quad \emph{Q.} Which of the following best describes who you are when using the healthcare chatbot?\quad \emph{A.} ``Users consulting on personal health issues.''\\[2pt]
\textbf{Privacy attitudes ($A_u$).}\quad \emph{Q.} Is online healthcare consultation more likely to compromise privacy than asking a doctor face-to-face?\quad \emph{A.} ``Somewhat agree'' (5-point: Strongly disagree~$\to$~Strongly agree).

\vspace{2pt}
\hrule height 0.3pt
\vspace{2pt}

\textbf{Data-sharing question $q_i$ (Q1).}\quad Imagine you use the AI-large-scale-model for healthcare consultation, and the conversation mentions your \emph{mental health information} (\eg stress levels, emotional fluctuations, psychotherapy experiences). How acceptable is it for \emph{your family members} to access this information?\\[2pt]
\textbf{User response $r_{u,i}$.}\quad ``Somewhat acceptable'' \;(5-point: \emph{completely unacceptable / somewhat unacceptable / neutral / somewhat acceptable / completely acceptable}; mapped to $\{0,1,2,3,4\}$ for evaluation).
\end{tcolorbox}

\subsection{AI Agent Permissions}
\paragraph{Context.} \citet{wu2025towards} ask participants to imagine training a future personal AI assistant (``MIRA'') in 2049 and to set permission preferences over (data, app) tuples that the assistant would request to fulfill a natural-language query (\eg trip planning, banking, fitness, home security, pregnancy support). Each item is answered on a four-point preference scale: \emph{Yes, always share}; \emph{Yes, but ask me next time}; \emph{No, but ask me next time}; \emph{No, never share}.
\paragraph{Participants.} 203 U.S.\ adult participants. The released demographics cover education, age range, and gender.
\paragraph{Persona facets.} $D_u$ contains education, age range, and gender. $E_u$ records self-rated familiarity with current AI tools (1--5), frequency of use (1--5), and free-text purpose of use. $A_u$ contains general trust in AI tools (1--5), the importance of privacy when using digital tools (1--5), free-text privacy concerns, and the three application domains the participant cares most about.
\paragraph{Key findings.} Permission preferences are highly heterogeneous across users \emph{and} highly context-dependent within a user: the same data type (\eg location) is approved for one app/intent but not another, and few items receive a uniform allow/deny verdict across the population. This motivates per-user, per-context permission inference rather than global defaults.
\paragraph{Data availability.} The raw user-study data is released by the original authors at \url{https://github.com/llm-platform-security/ai-agent-permissions} under the Creative Commons Attribution 4.0 International (CC BY 4.0) license.
\begin{tcolorbox}[
    enhanced,
    breakable,
    colback=gray!3,
    colframe=black!55,
    boxrule=0.5pt,
    arc=2pt,
    left=6pt,right=6pt,top=4pt,bottom=4pt,
    fonttitle=\bfseries\small,
    coltitle=black,
    colbacktitle=gray!12,
    title={Example: User from AI Agent Permissions \citep{wu2025towards}},
    fontupper=\small,
]
\textbf{Demographics ($D_u$).}\quad Age: \emph{45--54};\; Gender: \emph{Male};\; Education: \emph{Master's}.\\[2pt]
\textbf{Previous experiences ($E_u$).}\quad \emph{Q.} On a scale from 1 (Not at all familiar) to 5 (Very familiar), how familiar are you with existing AI tools, such as ChatGPT and Gemini, in terms of your knowledge and usage?\quad \emph{A.} ``3'' (\emph{Somewhat familiar}).\\[2pt]
\textbf{Privacy attitudes ($A_u$).}\quad \emph{Q.} On a scale from 1 (Not at all) to 5 (Full trust), how much do you trust these tools to not violate your privacy?\quad \emph{A.} ``3''.

\vspace{2pt}
\hrule height 0.3pt
\vspace{2pt}

\textbf{Data-sharing question $q_i$ (Q1).}\quad You ask your personal AI assistant: ``I want to lose weight. Can you create an at-home full-body workout plan for me?'' To address this request, your AI assistant wants to access \emph{Gender} from \emph{Fitness Tracking}. How willing are you to share this data?\\[2pt]
\textbf{User response $r_{u,i}$.}\quad ``Yes, always share'' \;(4-point: \emph{Yes, always share / Yes, but ask me next time / No, but ask me next time / No, never share}; mapped to $\{3,2,1,0\}$ for evaluation).
\end{tcolorbox}

\subsection{LLM Chatbot}
\paragraph{Context.} \citet{tran2025understanding} apply Helen Nissenbaum's contextual-integrity framework to ChatGPT-history sharing. Each scenario is a factorial vignette varying along five contextual-integrity attributes: \emph{recipient} (university CS/social-science researcher, big tech, insurance company, hospital, charitable foundation, government, among others), \emph{location} (US, EU, China), \emph{purpose} (assessing mental health, predicting behavior, training future AI models, advertising, dataset creation, fighting terrorism, AI-risk research), \emph{topic} of conversation (job, social/personal, legal/ethical, health, all), and three transmission principles (\emph{informed/consent/none}, \emph{anonymized or not}, \emph{personally identifiable information (PII) automatically removed or not}). Participants rate appropriateness on a 0--100 numeric scale.
\paragraph{Participants.} 300 U.S.\ ChatGPT users. Demographics are not released in the public dataset, so $D_u = \emptyset$.
\paragraph{Persona facets.} $E_u$ records ChatGPT use frequency, first-use date, subscription type, the tasks the participant uses ChatGPT for, and the topics they discuss with it (with optional free-text elaboration). $A_u$ records five 0--100 statements: companies collect too much data, general trust in websites, importance of privacy, ChatGPT-specific privacy concern, and belief that conversations remain private.
\paragraph{Key findings.} Appropriateness ratings vary sharply with the contextual-integrity attributes, especially recipient and purpose; consent and anonymization raise appropriateness but do not fully compensate for a sensitive recipient/purpose pairing; and stated privacy attitudes only weakly predict appropriateness ratings.
\paragraph{Data availability.} The raw user-study data is released by the original authors on Harvard Dataverse at \url{https://dataverse.harvard.edu/dataset.xhtml?persistentId=doi:10.7910/DVN/M6ABJ3}~\citep{DVN/M6ABJ3_2025} under the Creative Commons Zero (CC0 1.0) public-domain dedication.
\begin{tcolorbox}[
    enhanced,
    breakable,
    colback=gray!3,
    colframe=black!55,
    boxrule=0.5pt,
    arc=2pt,
    left=6pt,right=6pt,top=4pt,bottom=4pt,
    fonttitle=\bfseries\small,
    coltitle=black,
    colbacktitle=gray!12,
    title={Example: User from LLM Chatbot \citep{tran2025understanding}},
    fontupper=\small,
]
\textbf{Demographics ($D_u$).}\quad \texttt{Not provided.}\\[2pt]
\textbf{Previous experiences ($E_u$).}\quad \emph{Q.} How frequently do you use ChatGPT?\quad \emph{A.} ``Less than monthly.''\\[2pt]
\textbf{Privacy attitudes ($A_u$).}\quad \emph{Q.} Online companies collect too much personal information.\quad \emph{A.} ``70'' (0--100: Strongly disagree~$\to$~Strongly agree).

\vspace{2pt}
\hrule height 0.3pt
\vspace{2pt}

\textbf{Data-sharing question $q_i$ (Q1).}\quad Your ChatGPT conversations were shared with a big tech company based in the US for the purpose of fighting terrorism. All of your conversations were shared, you were informed that your data was collected, your conversations were not anonymized, and personally identifying information was not removed. Please rate your agreement with: ``This is an appropriate use of my ChatGPT data.''\\[2pt]
\textbf{User response $r_{u,i}$.}\quad ``0'' \;(0--100 numeric scale: 0 = Strongly disagree, 50 = Neutral, 100 = Strongly agree).
\end{tcolorbox}

\subsection{Chatbot Group Chat}
\paragraph{Context.} \citet{chou2026bot} probe user awareness of and concern about chatbots that operate inside group chats on instant-messaging (IM) platforms (Discord, Telegram, WhatsApp, and Slack). After eliciting beliefs about what messages and metadata such chatbots can access, the questionnaire asks the participant to rate agreement with seven statements covering privacy concern, behavior change (would they self-censor?), continued use, and protection-policy understanding, on a five-point Likert scale.
\paragraph{Participants.} 374 participants worldwide recruited via Prolific. Demographics include age range, gender, race/ethnicity, highest level of education, and a binary technical-work-experience flag.
\paragraph{Persona facets.} $D_u$ contains age, gender, race/ethnicity, education, and technical experience. $E_u$ describes the IM service the participant uses most, frequency of IM and chatbot use, types of group chats and chatbot purposes encountered, free-text descriptions of specific chatbots seen, and habits around checking for chatbot presence in private vs.\ public groups. $A_u$ records the participant's beliefs about what message content, message metadata, sender metadata, and group-level information chatbots can access on their chosen platform.
\paragraph{Key findings.} Many participants underestimate the breadth of data accessible to group-chat bots, especially metadata and access to deleted messages; awareness varies substantially across platforms; and even highly aware users typically keep using the chatbots, citing utility.
\paragraph{Data availability.} The raw user-study data is released by the original authors at \url{https://github.com/csienslab/bot-among-us} under the MIT License.
\begin{tcolorbox}[
    enhanced,
    breakable,
    colback=gray!3,
    colframe=black!55,
    boxrule=0.5pt,
    arc=2pt,
    left=6pt,right=6pt,top=4pt,bottom=4pt,
    fonttitle=\bfseries\small,
    coltitle=black,
    colbacktitle=gray!12,
    title={Example: User from Chatbot Group Chat \citep{chou2026bot}},
    fontupper=\small,
]
\textbf{Demographics ($D_u$).}\quad Age: \emph{25--34};\; Gender: \emph{Male};\; Education: \emph{Bachelor's degree};\; Race/ethnicity: \emph{White or Caucasian};\; Technical work experience: \emph{Yes}.\\[2pt]
\textbf{Previous experiences ($E_u$).}\quad \emph{Q.} Please select one instant messaging service you use frequently and have seen chatbots on.\quad \emph{A.} ``Discord.''\\[2pt]
\textbf{Privacy attitudes ($A_u$).}\quad \emph{Q.} What type of messages do you believe a chatbot on Discord has access to?\quad \emph{A.} ``All messages sent in the chat, including those that are later deleted.''

\vspace{2pt}
\hrule height 0.3pt
\vspace{2pt}

\textbf{Data-sharing question $q_i$ (Q1).}\quad Please rate your level of agreement with the following statement about chatbots in group chats: ``Chatbots from group chats with access to users' data as listed above would raise my privacy concerns.''\\[2pt]
\textbf{User response $r_{u,i}$.}\quad ``Neither agree nor disagree'' \;(5-point: \emph{Strongly disagree / Somewhat disagree / Neither agree nor disagree / Somewhat agree / Strongly agree}; mapped to $\{0,1,2,3,4\}$ for evaluation).
\end{tcolorbox}

\subsection{LLM Conversational Agents}
\paragraph{Context.} \citet{zufferey2025ai} measure how frequently participants share each of ten categories of personal data with general-purpose conversational agents such as ChatGPT and locally hosted LLMs. Categories cover \emph{identifiers and general information} (personal identifiers, account credentials, location/mobility, ethnicity/citizenship, criminal records) and \emph{standard of living and opinions} (family/relationship, wealth, job/education, religion, political views), each rated on a five-point frequency scale from \emph{I never shared} to \emph{I usually shared}.
\paragraph{Participants.} 422 participants were recruited internationally with a European skew. We focus on just the chatbot users since they answered questions involving frequency of data disclosure, and drop users with missing frequency responses, narrowing the participant pool to 277. Released demographics are intentionally light: age range and country/region.
\paragraph{Persona facets.} $D_u$ contains age and country/region. $E_u$ describes whether the participant runs a local LLM instance (and, if so, why), the chatbot services they use, frequency and average daily duration of chatbot interaction, devices used, and their typical purposes. $A_u$ contains the eight items of the Internet Users' Information Privacy Concerns (IUIPC) privacy-concern scale (collection, control, awareness) on a seven-point agreement scale.
\paragraph{Key findings.} Self-reported sharing frequency is low overall but varies sharply by category: job/education-related and general-information sharing are common, while credentials, criminal records, religion, and political views are rarely shared. Users running their own local LLMs report more frequent sharing across most categories, consistent with greater perceived control.
\paragraph{Data availability.} The raw user-study data is released by the original authors on the Open Science Framework at \url{https://osf.io/2vqws/} under the Creative Commons Attribution 4.0 International (CC BY 4.0) license.
\paragraph{Example user.} A representative user from this study is shown in Example~\ref{ex:user-example}.

\section{Additional Results}\label{appendix:additional_results}

\begin{figure}[t]
  \centering
  \includegraphics[width=\linewidth]{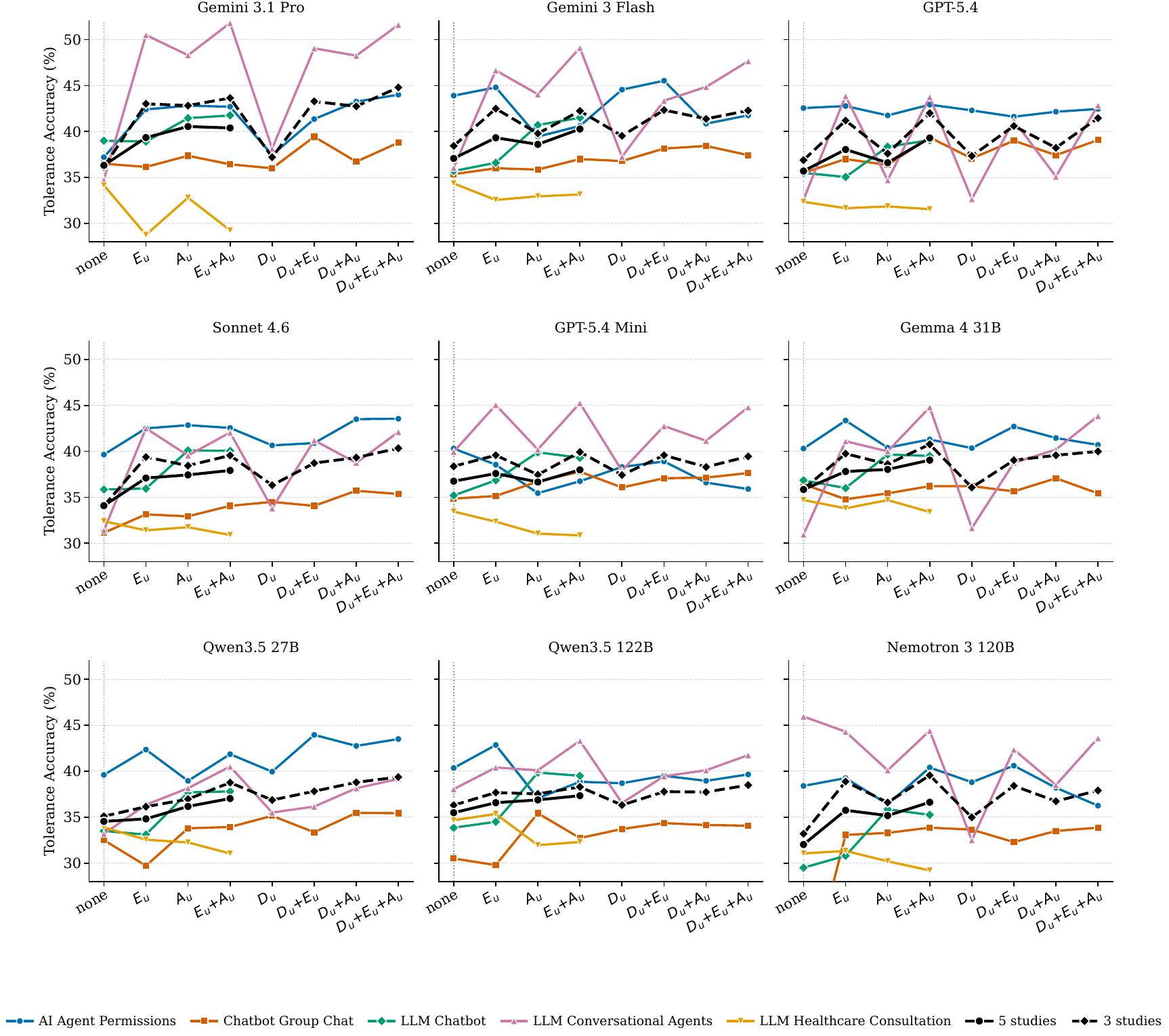}
  \caption{Per-domain tolerance accuracy across all eight prompt types for every model evaluated in \method{}, including Qwen3.5 122B, which is omitted from Figure~\ref{fig:main_results}. Solid black: 5-study average; dashed black: 3-study average. Each subplot shares the same y-axis so cross-model comparisons are direct.}
  \label{fig:domain_subplots_full}
\end{figure}

Figure~\ref{fig:domain_subplots} in the main paper shows per-domain accuracy for four closed-source models. Figure~\ref{fig:domain_subplots_full} below extends the results to all nine models evaluated in \method{} by adding GPT-5.4 Mini, Gemma 4 31B, Qwen3.5 27B, Qwen3.5 122B, and Nemotron 3 120B. 

We observe that across the nine models, LLM Conversational Agents (pink) sits at or near the top in every subplot, LLM Healthcare Consultation (yellow) sits at the bottom, and AI Agent Permissions (blue), Chatbot Group Chat (orange), and LLM Chatbot (green) cluster in the middle. We observe that the privacy paradox identified in Section~\ref{sec:experiments} is visible in many of the nine models. Conditioning on privacy attitudes alone reliably dips in performance when compared to conditioning on previous experiences alone. Lastly, we observe that conditioning on both privacy attitudes and previous experiences consistently improves performance across all nine models, confirming the main paper's conclusion that combining the two facets gives the most reliable improvement. 

\subsection{Majority-Class Baseline}\label{appendix:majority_baseline}

In this section, we include a non-LLM majority-class predictor to get a sense of how good the best LLM's accuracy of 40.4\% is. However, the two are not directly comparable because their setups differ. The majority-class predictor is fit on a training set of ground-truth responses from all participants in each source study, including the evaluated users' own responses, and for every user and question it predicts the most frequent response among the participants who answered the same question. In contrast, the LLM is evaluated zero-shot from a single user's privacy persona, which contains no ground-truth privacy decisions from any user, so it has no training set at all (Section~\ref{sec:problem-statement}). Despite this difference, the comparison is still useful because it shows how much of user privacy behavior is predictable from population-level privacy decisions alone. Table~\ref{tab:majority-baseline} reports the results. The majority-class predictor reaches 48.8\% averaged over the 5 studies, compared to 40.4\% for Gemini 3.1 Pro conditioned on $E_u{+}A_u$, which indicates that population-level privacy decisions, that is, general privacy norms, are a stronger predictor of user privacy behavior than the persona facets.

\begin{table}[h]
  \centering
  \caption{Tolerance accuracy of the per-question majority-class baseline compared to Gemini 3.1 Pro conditioned on $E_u{+}A_u$, per study and averaged over the 5 studies. The majority-class baseline predicts, for every user, the most frequent response among the participants of the source study who answered the same question.}
  \label{tab:majority-baseline}
  \small
  \begin{tabular}{lrr}
    \toprule
    Domain & Majority class & Gemini 3.1 Pro ($E_u{+}A_u$) \\
    \midrule
    AI Agent Permissions        & 54.00 & 42.70 \\
    Chatbot Group Chat          & 37.43 & 36.43 \\
    LLM Chatbot                 & 55.55 & 41.75 \\
    LLM Conversational Agents   & 54.50 & 51.80 \\
    LLM Healthcare Consultation & 42.50 & 29.25 \\
    \midrule
    Average over 5 studies      & 48.80 & 40.39 \\
    \bottomrule
  \end{tabular}
\end{table}

\subsection{Sweep of \texorpdfstring{$\tau$}{tau} for Tolerance Accuracy}\label{appendix:tau_sweep}

Table~\ref{tab:tau-sweep} sweeps $\tau$ for Gemini 3.1 Pro. Results change modestly for $\tau$ between $0$ and $0.20$, and the gain of $E_u{+}A_u$ over \texttt{none} stays at roughly 4--5\%, so our RQ1 conclusion does not depend on the choice of $\tau$ in this range. At $\tau=0$, however, accuracy on LLM Chatbot drops to 5--12\%, since responses on its 100-point scale rarely match exactly. At $\tau=0.25$ there is a sharp jump: the tolerance reaches one full category on the 5-point studies, so answers that are one category away (\eg ``Somewhat agree'' vs.\ ``Neutral'') suddenly count as correct. We choose $\tau=0.15$ because it is within the stable region while avoiding the sharp drop on LLM Chatbot.

\begin{table}[h]
  \centering
  \caption{Tolerance accuracy of Gemini 3.1 Pro as a function of $\tau$, for the no-persona prompt (\texttt{none}) and the $E_u{+}A_u$ prompt. The top block averages over all 5 studies; the bottom block is LLM Chatbot alone, the only study with a 0--100 scale. The column for our choice $\tau=0.15$ is in bold.}
  \label{tab:tau-sweep}
  \small
  \setlength{\tabcolsep}{5pt}
  \begin{tabular}{lrrrrrrrr}
    \toprule
    Prompt type & $\tau{=}0$ & $0.10$ & $\mathbf{0.15}$ & $0.20$ & $0.25$ & $0.50$ & $0.75$ & $1.00$ \\
    \midrule
    \multicolumn{9}{l}{\emph{Average over all 5 studies}} \\
    \texttt{none}  & 29.48 & 34.54 & \textbf{36.32} & 37.81 & 62.32 & 85.92 & 95.89 & 100.00 \\
    $E_u{+}A_u$    & 34.39 & 38.60 & \textbf{40.39} & 41.76 & 62.65 & 83.89 & 95.21 & 100.00 \\
    \midrule
    \multicolumn{9}{l}{\emph{LLM Chatbot only}} \\
    \texttt{none}  &  4.80 & 30.10 & \textbf{39.00} & 46.45 & 52.45 & 77.25 & 94.95 & 100.00 \\
    $E_u{+}A_u$    & 11.75 & 32.80 & \textbf{41.75} & 48.60 & 54.05 & 76.50 & 92.90 & 100.00 \\
    \bottomrule
  \end{tabular}
\end{table}

\section{Additional Details on Privacy Persona Cluster}\label{appendix:privacy_persona_cluster}\label{app:clustering}

Section~\ref{sec:rq2} groups every participant in \method{} into one of five privacy-persona clusters using a (stance, exposure) layout inspired by \citet{dupree2016privacy} and asks which clusters are hardest for an LLM to simulate. This appendix goes through which user features are used in the clustering procedure, how those features are standardized, how the five clusters are then assigned, and what design choices were made to keep the analysis honest.

Each user in \citet{dupree2016privacy} is placed on two axes: \emph{motivation} (how strongly they care about protecting their privacy) and \emph{knowledge} (how much technical understanding they have of privacy risks). Because the source studies in \method{} do not include a privacy-knowledge instrument, we cannot use this knowledge axis directly. We therefore borrow the two-axis layout and instantiate the axes from the persona facets we do have:
\begin{itemize}
    \item \textbf{Stated stance, $s_u$:} a $[0,1]$ score summarizing the participant's \emph{privacy attitudes} ($A_u$).
    \item \textbf{Prior exposure, $e_u$:} a $[0,1]$ score summarizing the participant's \emph{previous experiences} ($E_u$), capturing AI-tool familiarity rather than privacy knowledge.
\end{itemize}

Thus, we map responses under privacy attitudes and previous experiences into a score within $[0, 1]$. Privacy-attitude statements across the five studies use either 5-point agreement Likert, familiarity Likert, or numeric scores (1--5, 0--100), all of which are easy to map to $[0,1]$. However, a high Likert score can indicate either high or low privacy concern depending on how the statement is framed. We therefore parse each statement and decide its polarity by keyword: if it contains concern-direction words (\emph{concerned}, \emph{worried}, \emph{compromise}, $\ldots$), a high score indicates high concern and the raw $[0,1]$ value $x$ is kept; if it contains permissive-direction words (\emph{trust}, \emph{willing}, \emph{will use}, $\ldots$), a high score indicates low concern, so we flip it to $1-x$. After this correction, ``higher value = more concerned'' for every item, and $s_u$ is the simple mean over all polarity-corrected items.

Previous-experience statements use a wider mix of formats (frequency phrases, yes/no, numeric familiarity, free text, multi-select chips), so we parse each statement and decide its score by format. Frequency phrases map to a fractional value (\emph{daily} $\to 1$, \emph{often} $\to 0.75$, \emph{sometimes} $\to 0.5$, $\ldots$, \emph{never} $\to 0$); a yes / has-used / has-experience answer scores $1$; a numeric familiarity rating $\geq 3$ on a 1--5 scale scores $1$. Anything else, including a ``no'' answer, a low familiarity rating, free text, or unrecognized multi-select chips, scores $0$. We then take $e_u$ as the mean of these scores over all non-skipped items.

The five source studies use different response vocabularies and different distributions of stance and exposure (\eg the LLM Conversational Agents pool is heavily European and tech-comfortable; the LLM Healthcare pool skews toward general health users). We therefore z-score $s_u$ and $e_u$ \emph{within each domain} to obtain $\hat{s}_u$ and $\hat{e}_u$, so that ``high stance'' and ``high exposure'' have a consistent meaning regardless of which study a participant came from. Note that this step takes the raw scores out of $[0,1]$: $\hat{s}_u$ and $\hat{e}_u$ are real-valued, centered at $0$ with unit variance \emph{within each domain}, so a value of $0$ means ``average for this domain'', positive means above average, and negative means below average.

We place the five clusters at fixed canonical centers in $(\hat{e}_u, \hat{s}_u)$ space. Because $\hat{s}_u$ and $\hat{e}_u$ are z-scores, the natural ``high / mid / low'' tiers sit at $+0.8$, $0$, and $-0.8$, the midpoints of the upper, middle, and lower tertiles of a standard normal, which gives the following centers:

\begin{center}
\small
\begin{tabular}{lccl}
\toprule
Cluster (stance / exposure) & $\hat{s}$ center & $\hat{e}$ center & Closest \citet{dupree2016privacy} cluster \\
\midrule
high / high   & $+0.8$ & $+0.8$ & Fundamentalist \\
high / mid    & $+0.8$ & $\phantom{+}0.0$ & Technician \\
mid / mid     & $\phantom{+}0.0$ & $\phantom{+}0.0$ & Amateur \\
low / high    & $-0.8$ & $+0.8$ & Lazy Expert \\
low / low     & $-0.8$ & $-0.8$ & Marginally Concerned \\
\bottomrule
\end{tabular}
\end{center}

Every user is then assigned to the nearest centroid under squared Euclidean distance in $(\hat{e}_u, \hat{s}_u)$. We chose fixed canonical centers (rather than k-means) so that a cluster label has the same meaning across studies.

Across the 1{,}000 participants in \method{}, the resulting cluster sizes are high-stance / mid-exposure = 226, low-stance / high-exposure = 221, low-stance / low-exposure = 205, mid-stance / mid-exposure = 187, high-stance / high-exposure = 161. To check that the qualitative findings in Section~\ref{sec:rq2} are not artifacts of the specific center placement, we also re-ran the assignment with centers pushed to $\pm 0.30$ and $\pm 0.55$ (corresponding to the conservative and aggressive tertile choices, respectively). Under both perturbations, low-stance / high-exposure remains the lowest-accuracy cluster at $E_u{+}A_u$ and high-stance / mid-exposure remains the cluster that gains the most from persona conditioning.
\clearpage
\section*{NeurIPS Paper Checklist}

The checklist is designed to encourage best practices for responsible machine learning research, addressing issues of reproducibility, transparency, research ethics, and societal impact. Do not remove the checklist: {\bf The papers not including the checklist will be desk rejected.} The checklist should follow the references and follow the (optional) supplemental material.  The checklist does NOT count towards the page
limit. 

Please read the checklist guidelines carefully for information on how to answer these questions. For each question in the checklist:
\begin{itemize}
    \item You should answer \answerYes{}, \answerNo{}, or \answerNA{}.
    \item \answerNA{} means either that the question is Not Applicable for that particular paper or the relevant information is Not Available.
    \item Please provide a short (1--2 sentence) justification right after your answer (even for \answerNA). 
\end{itemize}

{\bf The checklist answers are an integral part of your paper submission.} They are visible to the reviewers, area chairs, senior area chairs, and ethics reviewers. You will also be asked to include it (after eventual revisions) with the final version of your paper, and its final version will be published with the paper.

The reviewers of your paper will be asked to use the checklist as one of the factors in their evaluation. While \answerYes{} is generally preferable to \answerNo{}, it is perfectly acceptable to answer \answerNo{} provided a proper justification is given (e.g., error bars are not reported because it would be too computationally expensive'' or ``we were unable to find the license for the dataset we used''). In general, answering \answerNo{} or \answerNA{} is not grounds for rejection. While the questions are phrased in a binary way, we acknowledge that the true answer is often more nuanced, so please just use your best judgment and write a justification to elaborate. All supporting evidence can appear either in the main paper or the supplemental material, provided in appendix. If you answer \answerYes{} to a question, in the justification please point to the section(s) where related material for the question can be found.

IMPORTANT, please:
\begin{itemize}
    \item {\bf Delete this instruction block, but keep the section heading ``NeurIPS Paper Checklist"},
    \item  {\bf Keep the checklist subsection headings, questions/answers and guidelines below.}
    \item {\bf Do not modify the questions and only use the provided macros for your answers}.
\end{itemize}


\begin{enumerate}

\item {\bf Claims}
    \item[] Question: Do the main claims made in the abstract and introduction accurately reflect the paper's contributions and scope?
    \item[] Answer: \answerTODO{} 
    \item[] Justification: \justificationTODO{}

\item {\bf Limitations}
    \item[] Question: Does the paper discuss the limitations of the work performed by the authors?
    \item[] Answer: \answerTODO{} 
    \item[] Justification: \justificationTODO{}

\item {\bf Theory assumptions and proofs}
    \item[] Question: For each theoretical result, does the paper provide the full set of assumptions and a complete (and correct) proof?
    \item[] Answer: \answerTODO{} 
    \item[] Justification: \justificationTODO{}

    \item {\bf Experimental result reproducibility}
    \item[] Question: Does the paper fully disclose all the information needed to reproduce the main experimental results of the paper to the extent that it affects the main claims and/or conclusions of the paper (regardless of whether the code and data are provided or not)?
    \item[] Answer: \answerTODO{} 
    \item[] Justification: \justificationTODO{}

\item {\bf Open access to data and code}
    \item[] Question: Does the paper provide open access to the data and code, with sufficient instructions to faithfully reproduce the main experimental results, as described in supplemental material?
    \item[] Answer: \answerTODO{} 
    \item[] Justification: \justificationTODO{}

\item {\bf Experimental setting/details}
    \item[] Question: Does the paper specify all the training and test details (e.g., data splits, hyperparameters, how they were chosen, type of optimizer) necessary to understand the results?
    \item[] Answer: \answerTODO{} 
    \item[] Justification: \justificationTODO{}

\item {\bf Experiment statistical significance}
    \item[] Question: Does the paper report error bars suitably and correctly defined or other appropriate information about the statistical significance of the experiments?
    \item[] Answer: \answerTODO{} 
    \item[] Justification: \justificationTODO{}

\item {\bf Experiments compute resources}
    \item[] Question: For each experiment, does the paper provide sufficient information on the computer resources (type of compute workers, memory, time of execution) needed to reproduce the experiments?
    \item[] Answer: \answerTODO{} 
    \item[] Justification: \justificationTODO{}
    
\item {\bf Code of ethics}
    \item[] Question: Does the research conducted in the paper conform, in every respect, with the NeurIPS Code of Ethics \url{https://neurips.cc/public/EthicsGuidelines}?
    \item[] Answer: \answerTODO{} 
    \item[] Justification: \justificationTODO{}

\item {\bf Broader impacts}
    \item[] Question: Does the paper discuss both potential positive societal impacts and negative societal impacts of the work performed?
    \item[] Answer: \answerTODO{} 
    \item[] Justification: \justificationTODO{}
    
\item {\bf Safeguards}
    \item[] Question: Does the paper describe safeguards that have been put in place for responsible release of data or models that have a high risk for misuse (e.g., pre-trained language models, image generators, or scraped datasets)?
    \item[] Answer: \answerTODO{} 
    \item[] Justification: \justificationTODO{}

\item {\bf Licenses for existing assets}
    \item[] Question: Are the creators or original owners of assets (e.g., code, data, models), used in the paper, properly credited and are the license and terms of use explicitly mentioned and properly respected?
    \item[] Answer: \answerTODO{} 
    \item[] Justification: \justificationTODO{}

\item {\bf New assets}
    \item[] Question: Are new assets introduced in the paper well documented and is the documentation provided alongside the assets?
    \item[] Answer: \answerTODO{} 
    \item[] Justification: \justificationTODO{}

\item {\bf Crowdsourcing and research with human subjects}
    \item[] Question: For crowdsourcing experiments and research with human subjects, does the paper include the full text of instructions given to participants and screenshots, if applicable, as well as details about compensation (if any)? 
    \item[] Answer: \answerTODO{} 
    \item[] Justification: \justificationTODO{}

\item {\bf Institutional review board (IRB) approvals or equivalent for research with human subjects}
    \item[] Question: Does the paper describe potential risks incurred by study participants, whether such risks were disclosed to the subjects, and whether Institutional Review Board (IRB) approvals (or an equivalent approval/review based on the requirements of your country or institution) were obtained?
    \item[] Answer: \answerTODO{} 
    \item[] Justification: \justificationTODO{}

\item {\bf Declaration of LLM usage}
    \item[] Question: Does the paper describe the usage of LLMs if it is an important, original, or non-standard component of the core methods in this research? Note that if the LLM is used only for writing, editing, or formatting purposes and does \emph{not} impact the core methodology, scientific rigor, or originality of the research, declaration is not required.
    \item[] Answer: \answerTODO{} 
    \item[] Justification: \justificationTODO{}

\end{enumerate}

\end{document}